\documentclass[reprint,aps,prc,twocolumn,amsmath,fleqn,dvipsnames,nofootinbib,superscriptaddress,doi=false]{revtex4-1}

\usepackage{tikz}
\usetikzlibrary{shapes,positioning} 

\usepackage[utf8]{inputenc}
\usepackage[T1]{fontenc}

\usepackage{pgfplots}
\pgfplotsset{compat=newest}

\usepackage{graphicx}
\usepackage{rotating}
\usepackage{pifont}
\usepackage{xcolor}

\usepackage{scalefnt}

\usepackage{amsmath}
\usepackage{amssymb}
\usepackage{amsfonts}
\usepackage{bm}
\usepackage{xfrac}
\usepackage{braket}

\usepackage{lmodern}
\usepackage{microtype}

\usepackage{scalefnt}

\newcommand{\norm}[1]{\left\lVert#1\right\rVert}

\pgfplotsset{compat=newest}

\definecolor{indianred}{rgb}{0.86, 0.08, 0.24}
\definecolor{royalblue}{rgb}{0.25, 0.41, 0.88}
\definecolor{darkorange}{rgb}{1.0, 0.55, 0}
\definecolor{mediumseagreen}{rgb}{0.24, 0.70, 0.44}
\definecolor{purple}{rgb}{0.5, 0, 0.5}
\definecolor{cyan3}{rgb}{0, 0.80, 0.80}

\definecolor{plot1}{rgb}{0.86, 0.08, 0.24}
\definecolor{plot2}{rgb}{0.25, 0.41, 0.88}
\definecolor{plot3}{rgb}{1.0, 0.55, 0}
\definecolor{plot4}{RGB}{61,153,86}

\newcommand{\la}{\langle}
\newcommand{\ra}{\rangle}

\newcommand{\Jtens}[4]{\ensuremath{{^{#2\hspace{-2pt}} #1}^{#3}_{#4}}}

\newcommand{\coll}[1]{\ensuremath{\mathbf{#1}}}

\begin{document}

\title{Tensor-decomposition techniques for \emph{ab initio} nuclear structure calculations. \\
From chiral nuclear potentials to ground-state energies}

\author{A. Tichai}
\affiliation{ESNT, CEA-Saclay, DRF, IRFU, D\'epartement de Physique Nucl\'eaire, Universit\'e de Paris Saclay, F-91191 Gif-sur-Yvette}
\email{alexander.tichai@cea.fr}

\author{R. Schutski}
\affiliation{Department of Chemistry, Rice University, Houston, TX 77005-1892}
\affiliation{CDISE, Skolkovo Institute of Science and technology, 3 Nobel st, Moscow, Russia}
\email{r.schutski@skoltech.ru}

\author{G. E. Scuseria}
\affiliation{Department of Chemistry, Rice University, Houston, TX 77005-1892}
\affiliation{Department of Physics and Astronomy, Rice University, Houston, TX 77005-1892}
\email{guscus@rice.edu}

\author{T. Duguet}
\affiliation{IRFU, CEA, Universit\'e Paris-Saclay, 91191 Gif-sur-Yvette, France}
\affiliation{KU Leuven, Instituut voor Kern- en Stralingsfysica, 3001 Leuven, Belgium}
\email{thomas.duguet@cea.fr}

\begin{abstract}
\begin{description}
\item[Background] The computational resources needed to generate the \emph{ab initio} solution of the nuclear many-body problem for increasing mass number and/or accuracy necessitates innovative developments to improve upon (1) the storage of many-body operators and (2) the scaling of many-body methods used to evaluate nuclear observables.  
The storing and efficient handling of many-body operators with high particle ranks is currently one of the major bottlenecks limiting the applicability range of \emph{ab initio} studies with respect to mass number and accuracy. 
Recently, the application of tensor decomposition techniques to many-body tensors has proven highly beneficial to reduce the computational cost of \emph{ab initio} calculations in quantum chemistry and solid-state physics.
\item[Purpose] The impact of applying state-of-the-art tensor factorization techniques to modern nuclear Hamiltonians derived from chiral effective field theory is investigated. Subsequently, the error induced by the tensor decomposition of the input Hamiltonian on ground-state energies of closed-shell nuclei calculated via second-order many-body perturbation theory is benchmarked.
\item[Methods] The first proof-of-principle application of tensor-decomposition techniques to the nuclear Hamiltonian is performed. Two different tensor formats are investigated by systematically benchmarking the approximation error on matrix elements stored in various bases of interest. The analysis is achieved while including normal-ordered three-nucleon interactions that are nowadays used as input to the most advanced \emph{ab initio} calculations in medium-mass nuclei.
With the aid of the factorized Hamiltonian, the second-order perturbative correction to ground-state energies is decomposed and the scaling properties of the underlying tensor network are discussed.
\item[Results] The employed tensor formats are found to lead to an efficient data compression of two-body matrix elements of the nuclear Hamiltonian. In particular, the sophisticated \emph{tensor hypercontraction} (THC) scheme yields low tensor ranks with respect to both harmonic-oscillator and Hartree-Fock single-particle bases. It is found that the tensor rank depends on the two-body total angular momentum $J$ for which one performs the decomposition, which is itself directly related to the sparsity the corresponding tensor. Furthermore, including normal-ordered two-body contributions originating from three-body interactions does not compromise the efficient data compression. Ultimately, the use of factorized matrix elements authorizes controlled approximations of the exact second-order ground-state energy corrections. In particular, a small enough error is obtained from low-rank factorizations in $^{4}$He, $^{16}$O and $^{40}$Ca.
\item[Conclusions] It is presently demonstrated that tensor-decomposition techniques can be efficiently applied to systematically approximate the nuclear many-body Hamiltonian in terms of lower-rank tensors. Beyond the input Hamiltonian, tensor-decomposition techniques can be envisioned to scale down the cost of state-of-the-art non-perturbative many-body methods in order to extend \emph{ab initio} studies to (i) higher precisions, (ii) larger masses and (iii) nuclei of doubly open-shell character.
\end{description}
\vspace{3pt}

\end{abstract}


\maketitle

\section{Introduction}

The processing of large data objects has become increasingly important in various fields of scientific activities.
In particular, high-dimensional data arrays denoted as tensors appear naturally in (nuclear) many-body theory, e.g., as matrix elements of the Hamiltonian. Dealing with larger and larger systems typically requires to increase the dimension of the $k$-body basis used to represent the $k$-body operators at play. Doing so, the required memory grows polynomially in basis size and exponentially in $k$, making the storage of large tensors eventually computationally demanding if not unfeasible. In order to overcome the storage problem of such tensors, efficient tensor-decomposition formats have been proposed and implemented, thus, allowing for more economic representations while controlling the approximation error.

Numerous applications in quantum chemistry and solid-state physics have shown that the electron repulsion tensor governing the Coulomb interaction between electrons can be efficiently decomposed using various tensor formats such as polyadic decomposition, density fitting or tensor hypercontraction~\cite{KoBa09,Ho12a,Ho12b,Pa12,Parrish13,Schu17,Hu17,bene11mp2, bene13ccd,MoSh18}.
While this leads to significant savings from a storage point of view, another major advantage of decomposing the interaction tensor can be obtained by reformulating many-body theories such that they display lower computational scaling.
For example it was realized that tensor decompositions yield versions of many-body perturbation theory (MBPT)~\cite{Ho12a,khoromskaia14a} and coupled cluster (CC) theory~\cite{Ho12a,Schu17} that have lower computational complexity than their original counterparts. In this way electronic correlations can be efficiently captured even for large molecular systems up to chemical accuracy. It is our objective to apply the same rationale to the nuclear many-body problem.

The nuclear many-body problem differs in two main aspects from its electronic counterpart.
(1) The size of the underlying one-body basis necessary to reach model-space convergence is significantly larger. 
(2) The inclusion of (at least) three-body operators in the Hamiltonian is mandatory for an accurate description of nuclear phenomena,
\begin{align}
H = T_\text{int}+ V + W + ... \, ,
\end{align}
where $T_\text{int}$ is the two-body intrinsic kinetic energy operator whereas  $V$ and $W$ denote two-nucleon (2N) and three-nucleon (3N) interactions, respectively. In particular, the storage of matrix elements of $W$ is currently a limiting factor to perform \emph{ab initio} calculations of heavy nuclei. It is worth noting that, while only allowing for proof-of-principle nuclear calculations, tensors encountered in the present publication have comparable dimension to those used in the most advanced quantum chemistry applications.

In this context, four major components of a full-fledged application of tensor-decomposition techniques can be envisioned in nuclear theory
\begin{enumerate}
\item The memory requirement of mode-6 tensors associated with three-nucleon forces currently constitutes a bottleneck for converged ab initio calculations of nuclei with mass number $A > 100$. Factorizing the initial tensor expressed in a large enough basis of reference prior to porting it to high-performance computing facilities used to perform many-body calculations (with or without resorting to the so-called normal-ordered two-body approximation) constitutes a long-term challenge. Achieving this task will require to develop and implement efficient factorization formats for mode-6 tensors~\cite{Lathauwer00a}, which is currently not as mature as for mode-4 tensors. Having efficient tensor formats for mode-6 tensors will be useful for other developments (see point 2.b below).
\item In the meantime, the tensor factorization of two-body operators can 
\begin{enumerate}
\item lower the memory requirements. In particular, there is strong interest in extending \emph{ab initio} nuclear calculations to (doubly) open-shell nuclei far away from shell closures~\cite{SoCi13,Hergert:2014iaa,Ja14,Si15,Bo14,H15,Geb17,Ti18a,Ti18b}.
When resorting to a single-reference approach, the treatment of such systems requires the spontaneous breaking~\cite{SoCi13,Ti18b,Art18a} (and restoration~\cite{Du15,Du16,Qiu17}) of symmetries, most importantly rotational ($SU(2)$) and global-gauge ($U(1)$) symmetries. From the practical viewpoint, the breaking of symmetries requires to work with a symmetry-broken basis. Using a $m$-scheme (when breaking $SU(2)$) and/or a quasiparticle (when breaking $U(1)$) formulation, the gain from using efficient factorizations of mode-4 (or more) tensors is expected to be particularly significant.
\item decrease the computational scaling of state-of-the art many-body calculations.  This is achieved by optimizing the contraction pattern between the Hamiltonian tensor and many-body tensors at play in the particular framework of interest. In order to do so, many-body formalisms must be reformulated from the outset in terms of decomposed tensors. In many-body perturbation theory, the many-body tensors are known \textit{a priori} (and analytically) and can be efficiently factorized as illustrated in the present paper. In non-perturbative methods, the many-body tensors in question constitute the unkowns to be solved for, e.g., the cluster amplitudes in CC theory, the transformed Hamiltonian generated through the in-medium similarity renormalization group (IMSRG) flow~\cite{Hergert:2017kx} or the coupling and interaction matrices in the ADC(n) implementation of self-consistent Green's function (SCGF) theory~\cite{Dickhoff:2004xx,Raimondi:2018aa}. By making an ansatz for the individual tensors, one can reformulate the many-body formalism to eventually solve \emph{for the factors in a specific tensor format} rather than for the full tensor; see Ref.~\cite{Schu17} for CC theory. The versatility of the employed tensor format allows for efficient tensor contractions, thus yielding lower computational scaling than the original (i.e. non-decomposed) framework. In order to fully exploit this idea, decomposing mode-6, e.g. CC triple amplitudes, tensors is eventually necessary.
\end{enumerate}
\item Ultimately, it is desirable to completely avoid the construction of the full mode-$n$ tensors entering the nuclear Hamiltonian and rather produce a factorized ansatz \emph{from the outset}, i.e., instead of performing an expensive decomposition of the nuclear Hamiltonian, the \emph{direct construction of tensor-factorized chiral matrix elements} is envisioned. Correspondingly, a reformulation of the renormalization group transformation of the Hamiltonian can be anticipated in terms of the tensor factors only, e.g., by deriving (free-space) similarity renormalization group flow equations for the factors rather than for the full tensor.
\end{enumerate}

We believe that the development and extension of tensor-decomposition techniques can significantly help pushing back the \emph{curse of dimensionality} of nuclear many-body theories and its individual manifestation at each of the aforementioned stages. While this first publication is dedicated to initiating point (2), the development of the various points elaborated on above requires steady formal and numerical developments in the years to come.

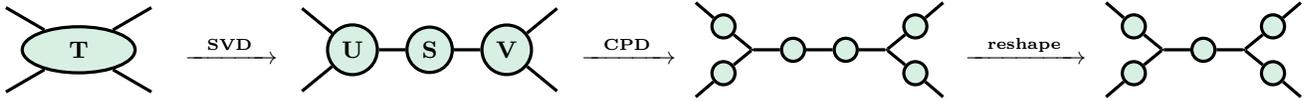
\begin{figure*}[t]
\begin{tikzpicture}
\newcommand{\lwidth}{1.2pt}
\newcommand{\edh}{12pt}
\newcommand{\edv}{8pt}
\newcommand{\edhcpd}{7pt}
\newcommand{\edvcpd}{5pt}
\newcommand{\nad}{15pt}
\newcommand{\innd}{10pt}
\newcommand{\arrsp}{6pt}
\node[ellipse,draw,line width=\lwidth,minimum width=1.5cm,fill=mediumseagreen!20] (t1) at (0,0) {$\mathbf{T}$};
\node[below left =\edv and \edh of t1] (u1) {};
\node[below right =\edv and \edh of t1] (u2) {};
\node[above left =\edv and \edh of t1] (l1) {};
\node[above right =\edv and \edh of t1] (l2) {};
\draw[line width=\lwidth] (t1) -- (u1);
\draw[line width=\lwidth] (t1) -- (u2);
\draw[line width=\lwidth] (t1) -- (l1);
\draw[line width=\lwidth] (t1) -- (l2);
\node[right=\nad of t1] (arrow1) {$\xrightarrow{\hspace{\arrsp} \textbf{SVD} \hspace{\arrsp} \vspace{5pt}}$};
\node[circle,draw,line width=\lwidth,fill=mediumseagreen!20,right =\nad of arrow1] (t2a) {$\mathbf{U}$};
\node[circle,draw,line width=\lwidth,fill=mediumseagreen!20,right =\innd of t2a] (t2b) {$\mathbf{S}$};
\node[circle,draw,line width=\lwidth,fill=mediumseagreen!20,right =\innd of t2b] (t2c) {$\mathbf{V}$};
\node[below left =\edv and \edh of t2a] (u1) {};
\node[above left =\edv and \edh of t2a] (l1) {};
\node[below right =\edv and \edh of t2c] (u2) {};
\node[above right =\edv and \edh of t2c] (l2) {};
\draw[line width=\lwidth] (t2a) -- (u1);
\draw[line width=\lwidth] (t2a) -- (l1);
\draw[line width=\lwidth] (t2c) -- (u2);
\draw[line width=\lwidth] (t2c) -- (l2);
\draw[line width=\lwidth] (t2a) -- (t2b);
\draw[line width=\lwidth] (t2b) -- (t2c);
\node[right=\nad of t2c] (arrow2) {$\xrightarrow{\hspace{\arrsp} \textbf{CPD} \hspace{\arrsp} \vspace{5pt}}$};
\node[circle,fill,inner sep=0pt,outer sep=0pt,right =\nad+10pt of arrow2] (t3a) {};
\node[circle,draw,line width=\lwidth,fill=mediumseagreen!20,right =\innd of t3a] (t3b) {};
\node[circle,draw,line width=\lwidth,fill=mediumseagreen!20,right =\innd of t3b] (t3c) {};
\node[circle,fill,inner sep=0pt,outer sep=0pt,right =\innd of t3c] (t3d) {};
\node[circle,draw,line width=\lwidth,fill=mediumseagreen!20,below left =\edvcpd and \edhcpd of t3a] (z1) {};
\node[circle,draw,line width=\lwidth,fill=mediumseagreen!20,above left =\edvcpd and \edhcpd of t3a] (z2) {};
\node[circle,draw,line width=\lwidth,fill=mediumseagreen!20,below right =\edvcpd and \edhcpd of t3d] (z3) {};
\node[circle,draw,line width=\lwidth,fill=mediumseagreen!20,above right =\edvcpd and \edhcpd of t3d] (z4) {};
\node[below left =\edvcpd and \edhcpd of z1] (l1) {};
\node[above left =\edvcpd and \edhcpd of z2] (u1) {};
\node[below right =\edvcpd and \edhcpd of z3] (l2) {};
\node[above right =\edvcpd and \edhcpd of z4] (u2) {};
\draw[line width=\lwidth] (t3a) -- (z1);
\draw[line width=\lwidth] (t3a) -- (z2);
\draw[line width=\lwidth] (t3d) -- (z3);
\draw[line width=\lwidth] (t3d) -- (z4);
\draw[line width=\lwidth] (z1) -- (l1);
\draw[line width=\lwidth] (z2) -- (u1);
\draw[line width=\lwidth] (z3) -- (l2);
\draw[line width=\lwidth] (z4) -- (u2);
\draw[line width=\lwidth] (t3a) -- (t3b);
\draw[line width=\lwidth] (t3b) -- (t3c);
\draw[line width=\lwidth] (t3c) -- (t3d);
\node[right=\nad+22pt of t3c] (arrow3){$\xrightarrow{\hspace{\arrsp} \textbf{reshape} \hspace{\arrsp} \vspace{5pt}}$};
\node[circle,fill,inner sep=0pt,outer sep=0pt,right =\nad+10pt of arrow3] (t4a) {};
\node[circle,draw,line width=\lwidth,fill=mediumseagreen!20,right =\innd of t4a] (t4b) {};
\node[circle,fill,inner sep=0pt,outer sep=0pt,right =\innd of t4b] (t4c) {};
\node[circle,draw,line width=\lwidth,fill=mediumseagreen!20,below left =\edvcpd and \edhcpd of t4a] (z1) {};
\node[circle,draw,line width=\lwidth,fill=mediumseagreen!20,above left =\edvcpd and \edhcpd of t4a] (z2) {};
\node[circle,draw,line width=\lwidth,fill=mediumseagreen!20,below right =\edvcpd and \edhcpd of t4c] (z3) {};
\node[circle,draw,line width=\lwidth,fill=mediumseagreen!20,above right =\edvcpd and \edhcpd of t4c] (z4) {};
\node[below left =\edvcpd and \edhcpd of z1] (l1) {};
\node[above left =\edvcpd and \edhcpd of z2] (u1) {};
\node[below right =\edvcpd and \edhcpd of z3] (l2) {};
\node[above right =\edvcpd and \edhcpd of z4] (u2) {};
\draw[line width=\lwidth] (t4a) -- (z1);
\draw[line width=\lwidth] (t4a) -- (z2);
\draw[line width=\lwidth] (t4c) -- (z3);
\draw[line width=\lwidth] (t4c) -- (z4);
\draw[line width=\lwidth] (z1) -- (l1);
\draw[line width=\lwidth] (z2) -- (u1);
\draw[line width=\lwidth] (z3) -- (l2);
\draw[line width=\lwidth] (z4) -- (u2);
\draw[line width=\lwidth] (t4a) -- (t4b);
\draw[line width=\lwidth] (t4b) -- (t4c);
\end{tikzpicture}
\label{fig:thcdiag}
\caption{(Color online) Diagrammatic representation of the THC decomposition of a mode-4 tensor.}
\end{figure*}

The present paper is organized as follows. Section \ref{sec:tensor} introduces the basic terminology of tensor calculus and present the tensor formats employed in this work. Section \ref{sec:num} presents the numerical calculations applied to a particular nuclear Hamiltonian. In Sec.~\ref{sec:gse}, the second-order correction to ground-state energies of selected doubly closed-shell nuclei is derived in its tensor-decomposed form and numerical results are displayed in Sec.~\ref{sec:mp2res}. Finally, conclusions are provided along with an outlook on future investigations in Sec.~\ref{sec:conclusion}.

\section{Tensor decompositions}
\label{sec:tensor}

\subsection{Basic terminology}

A \emph{mode-$n$ tensor} $T$ is defined as a $n$-dimensional multivariate data array $T_{i_1...i_n} $ with corresponding index ranges $\{I_1,...,I_n\}$\footnote{In the following one-body indices are denoted by lower-case letters and their index ranges by upper-case letters. Contrary multi-body indices and index ranges are denoted by lower- and upper-case bold letters, respectively.}. Except if stated otherwise, all presently studied tensors are such that $I_k = N$ for all $k=1,\ldots,n$. Special cases are vectors as mode-1 tensors and matrices as mode-2 tensors.
Therefore, high-order tensors can be seen as high-dimensional extensions of matrices.

The \emph{Frobenius norm} of a tensor $T$ is conveniently defined by
\begin{align}
\norm{ T } \equiv  \sqrt{\sum_{i_1...i_n} T_{i_1 ... i_n} T^\star_{i_1 ... i_n}}\, ,
\end{align}
where the superscript denotes elementwise complex conjugation.
Furthermore, the \emph{relative error} of the approximation $\hat T$ of a tensor $T$ is defined as
\begin{align}
\Delta T \equiv \frac{ \norm{ T - \hat T }}{\norm{ T}} \, .
\end{align}
In many cases the tensors under consideration are sparse, i.e., a large part of their entries are exactly zero.
This is typically related to symmetries of the object, e.g., constraints on single-particle angular-momentum, parity or isospin quantum numbers associated to the members of an index tuple.

\subsection{Tensor formats}

In this work two different ways of decomposing a tensor are investigated. 
A particular decomposition is referred to as a \emph{tensor format}. 
In the following the focus is entirely on mode-4 tensors.

The simplest factorization consists in performing a \emph{canonical polyadic decomposition} (CPD)~\cite{KoBa09} under the form
\begin{align}
T_{i_1 i_2 i_3 i_4}  = \sum_{\alpha} X^1_{i_1 \alpha} X^2_{i_2 \alpha} X^3_{i_3 \alpha} X^4_{i_4 \alpha}\, ,
\end{align}
which involves four factor matrices $\{X^i\}_{1\leq i \leq4}$. 
Given a tensor format, the size of the auxiliary summation\footnote{In the remainder auxiliary summation indices are denoted by greek indices.} index is referred to as the \emph{tensor rank}\footnote{It is worth noting that the notion of rank is in general not as strictly defined as for matrices, i.e., the rank concept is not unique for mode-$n$ tensors with $n\geq 3$.} 

For matrices, classical matrix factorizations such as QR- or LU-decompositions can be seen as CPD of matrices.
However, for $n\geq3$ no closed-form algorithm is known and one relies on an iterative solution to determine the factors.
Numerically, the factor matrices can be conveniently calculated using least-square minimization techniques, either of alternating least-square (ALS) or nonlinear least-square (NLS) type. The CPD is numerically challenging and the calculation of an $r_\text{CPD}$ approximation of a mode-$d$ tensor scales as
\begin{align}
\mathcal{O}(N^{d-1} r_\text{CPD} n_\text{iter} )\, ,
\end{align}
where $n_\text{iter}$ denotes the number of iterations to obtain the required accuracy and $r_\text{CPD}$ the required rank.

The CPD format was successfully applied in numerous quantum chemistry applications, in particular to compress the electron repulsion tensor. Empirically, it was found that the CPD rank necessary to reach chemical accuracy in many-body calculations, e.g., MBPT or CC,  scales as $N^{2.5}$~\cite{bene11mp2, bene13ccd}. The required storage for CPD scales as $4 N r_\text{CPD}$ which is less than the $N^4$ scaling of the initial tensor provided the CPD rank $r_\text{CPD}$ scales less than $\sim N^3$. 

A more sophisticated alternative is given by the so-called \emph{tensor hypercontraction} (THC) format~\cite{Ho12a,Ho12b,Schu17},
\begin{align}
T_{i_1 i_2 i_3 i_4}  = \sum_{\alpha \beta} X^1_{i_1 \alpha} X^2_{i_2 \alpha} W_{\alpha \beta} X^3_{i_3 \beta} X^4_{i_4 \beta} 
\label{eq:thc}
\end{align}
requiring five factor matrices, one of which is the core tensor $W$ that scales quadratically with the THC rank $r_\text{THC}$. In principle the index ranges of $\alpha$ and $\beta$ do not need to coincide even though it is always taken to be the case in our investigations. 

In practice the THC decomposition is realized via a multi-step approach: starting from a mode-4 tensor, the first and second pair of indices $\coll{k}=(i_1,i_2)$ and $\coll{l}=(i_3,i_4)$ are reshaped to obtain a matrix $T_{\coll{kl}}$ of size $\coll{K}\times \coll{L} = I_1I_2 \times I_3 I_4$. Subsequently, a singular value decomposition (SVD) is performed
\begin{align}
T_{\coll{k} \coll{l}} = \sum_{p} U_{\coll{k}  p} S_{p} V_{p\coll{l}}^T\, ,
\end{align}
where the (ordered) set of singular values is truncated, thus introducing the SVD rank $r_\text{SVD}$.
Next the diagonal matrix of singular values $S$ is absorbed into $U$ and $V$ by rewriting
\begin{subequations}
\begin{align}
\tilde U_{\coll{k}p} = U_{\coll{k} p} \sqrt{S_{p}} \, , \\
\tilde V^T_{p \coll{l}} =  \sqrt{S_{p}} V^T_{p \coll{l} }\, .
\end{align}
\label{eq:cpdthc}
\end{subequations}
Re-expanding the collective indices yields two mode-3 tensors
$\tilde U_{i_1 i_2 q}, \tilde V_{i_3 i_4 q}$
of size\footnote{This is the one case in our present study where all the indices of a given tensor do {\it not} have the same range.} $I_1 \times I_2 \times r_\text{SVD}$ and $I_3 \times I_4 \times r_\text{SVD}$, respectively.
The (rescaled) left and right factors are then used as input for a mode-3 CPD decomposition,
\begin{subequations}
\label{mode3CPD}
\begin{align}
\tilde U_{i_1 i_2 p} = \sum_{\alpha} X^1_{i_1 \alpha} X^2_{i_2 \alpha} Z^1_{p \alpha} \, , \\
\tilde V_{i_3 i_4 p} = \sum_{\beta} X^3_{i_3 \beta} X^4_{i_4 \beta} Z^2_{p \beta}\, .
\end{align}
\end{subequations}
In the last step, computing the core tensor through
\begin{align}
W_{\alpha \beta} \equiv \sum_p Z^1_{\alpha p} Z^2_{p \beta} 
\label{eq:core}
\end{align}
finalizes the individual operations of the THC decomposition in~\eqref{eq:thc}. For a graphical representation of the THC process, see Fig.~\ref{fig:thcdiag}. It is clear from Eq.~\eqref{eq:core} that the THC ranks $r_\text{THC}$ must not scale worse than $\sim N^2$. Otherwise, the memory required to store the core tensor exceeds the memory required for the initial tensor and the THC decomposition becomes superfluous.

\section{Nuclear Hamiltonian}
\label{sec:num}

The present section provides systematic decompositions of the nuclear Hamiltonian. While the THC was implemented in an in-house code suite, use was made of the Tensorlab library~\cite{tensorlab3.0} and, especially, of their high-level routines for the calculation of the CPD factors.

As a benchmark, a chiral effective field theory ($\chi$EFT) Hamiltonian containing 2N plus 3N interactions is employed after further softening it via a similarity renormalization group (SRG) transformation~\cite{BoFu07,HeRo07,RoRe08,RoLa11,JuMa13}.
The 2N interaction is constructed at next-to-next-to-next-to leading order in Weinberg's power counting with a cutoff $\Lambda_{2N}=500 \,\text{MeV}$~\cite{EnMa03} while the 3N force is constructed at next-to-next-to leading order with a cutoff $\Lambda_{3N}=400\,\text{MeV}$~\cite{Na07,Roth:2011vt}. Such a $\chi$EFT Hamiltonian has been routinely used in state-of-the-art many-body calculations~\cite{HeBi13,BiLa13,Ti16,Ti18a,Ti18b}. 

\subsection{$JT$-coupled matrix elements in the HO basis}

\begin{figure}
\centering
\includegraphics[width=0.9\columnwidth]{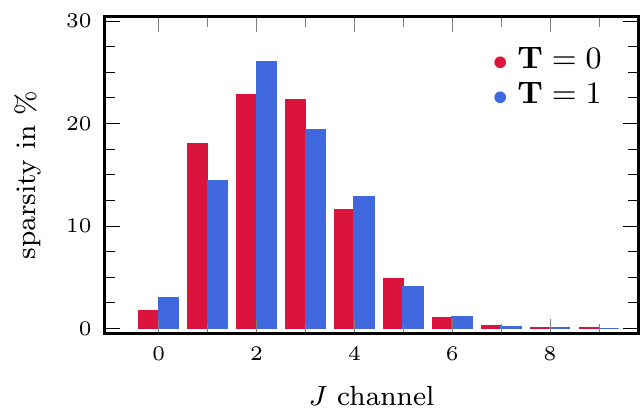}
\caption{(Color online) Histogram of the percentage of non-zero entries of the two-body nuclear interaction tensor for different $(J,T)$ channels in a $e_\text{max}=4$-truncated model space corresponding to a one-body index size $N=15$. The sparsity for different values of $M_T$ in the $T=1$ channel is the same since only the values of entries of the tensors are affected but not their non-zero character. }
\label{fig:sparsityJT}
\end{figure}

The first test is dedicated to matrix elements of the 2N interaction represented in the $JT$-coupled two-body basis built from a single-particle basis consisting of eigenfunctions of the three-dimensional spherical harmonic oscillator (HO) hamiltonian.
Therefore, the tensor is characterized by external block parameters $(J,T,M_T)$\footnote{Since $V$ is independent of the projection $M_J$ it is left out for convenience. Since the employed 2N interaction includes electromagnetic forces isospin symmetry is broken, such that the blocks do carry an explicit $M_T$ label.}, 
\begin{align}
V^{JTM_T}_{\breve \imath_1 \breve \imath_2 \breve \imath_3 \breve \imath_4} \equiv \la \breve \imath_1 \breve \imath_2 (JT M_T) | V | \breve \imath_3 \breve \imath_4 (JT M_T ) \ra\, ,
\label{eq:inttensJT}
\end{align}
where $J$ denotes the two-body total angular momentum, T the two-body isospin and $M_T$ its third component.
We subsequently refer to the object defined in~\eqref{eq:inttensJT} as the ($JT$-coupled) \emph{2N interaction tensor}.
Correspondingly, the indices of the tensor denote a reduced set of single-particle quantum numbers\footnote{Since the set~\eqref{eq:jt} neither contains total angular-momentum projection $m_i$ nor isospin projection $t_i$ it does not correspond to an actual single-particle state of the one-body Hilbert space $\mathcal{H}_1$. However, since we are presently working within a $JT$-coupled representation this set does fully specify the required information in each block.}
\begin{align}
\breve \imath \equiv (n_i, l_i, j_i) \, ,
\label{eq:jt}
\end{align}
where $n$ denotes the radial HO quantum number, $l$ the orbital angular-momentum quantum number and $j$ the total angular-momentum quantum number. The range $N$ of the interaction tensor indices is governed by the value of the principal quantum number $e_\text{max} = 2n +l$ that characterizes the dimension of the one-body Hilbert space used to represent the tensor. Accordingly, the maximum value of the two-body angular-momentum is given by $J_\text{max}= 2e_\text{max}+1$. In the present paper, all calculations are performed in a $e_\text{max} = 4$ single-particle space.

For future reference Fig.~\ref{fig:sparsityJT} depicts the percentage of non-zero matrix elements in the individual $(J,T,M_T)$ blocks. The significantly lower number of non-zero entries for extremal values of $J$ is linked to the triangular inequality for Clebsch-Gordan coefficients
\begin{align}
|j_{k_1} - j_{k_2} | \leq J \leq |j_{k_1} + j_{k_2} | \, ,
\end{align}
for the coupling of two single-particle angular momenta to a two-body angular momentum $J$.
For example, it imposes $j_{k_1}=j_{k_2}$ for $J=0$ and $j_{k_1}=j_{k_2}=l_\text{max} + \frac{1}{2}$\, where $l_\text{max}$ is the maximal orbital angular-momentum quantum number, for $J=J_\text{max}$.

\subsubsection{Canonical polyadic decomposition}

\begin{figure}[t!]
\includegraphics{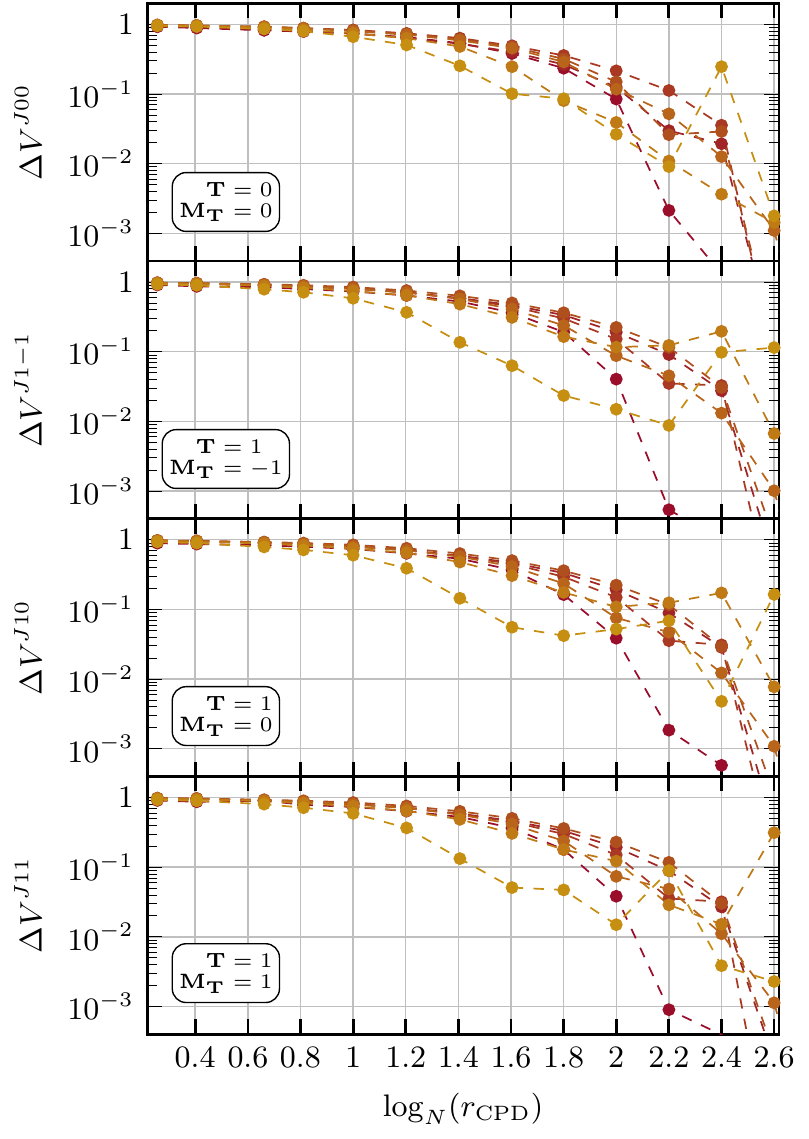}
\caption{(Color online) Relative error of the CPD format of $V^{JTM_T}$ for different $(J,T,M_T)$ blocks. Calculations are performed in a $e_\text{max}=4$ single-particle space. In each panel, increasing values of $J$ correspond to decreasingly darker curves.}
\label{fig:cpd}
\end{figure}

The CPD is applied by using NLS minimization to solve for the factor matrices. The number of maximum iterations used for the CPD algorithm is set to 600. For data compression it is key to quantify the required CPD rank $r_\text{CPD}$ needed to reach a given accuracy. Figure~\ref{fig:cpd} provides an overview of the results for different $(J,T,M_T)$ blocks.

The CPD rank exhibits a dependence on $J$ while being almost independent of $T$ and $M_T$.
The required rank is maximal for intermediate values of $J$. This observation is directly linked to the sparsity of the Hamiltonian (see Fig.~\ref{fig:sparsityJT}), i.e. the more non-zero entries\footnote{Note that the electron repulsion tensor is less dense than the $JT$-coupled nuclear interaction tensor.} in the original tensor, the higher the required CPD rank. Reaching a relative error\footnote{A relative error $\Delta V^{JTM_T}=10^{-1}$ typically corresponds to a deviation of $10^{-1}$ at the level of individual tensor entries.} of $\Delta V^{JTM_T} = 10^{-1}$ requires, in the worst-case scenario, $r_\text{CPD} \approx N^{2.5}$ for $J=2$. For other values of $J$ the required ranks scale much more mildly with the tensor range. In none of the cases the required rank exceeds $N^3$, which is the critical scaling for having an efficient representation of the interaction tensor in the CPD format. 

\begin{figure}[t!]
\centering
\includegraphics{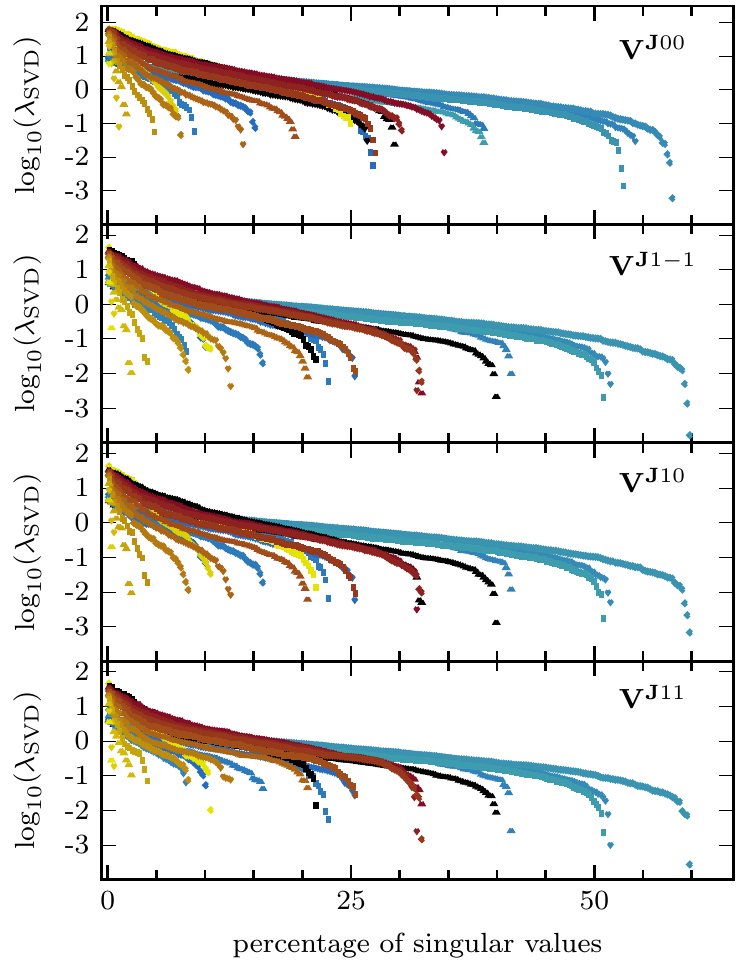}
\caption{(Color online) Size of the singular values extracted for $V^{JTM_T}$ for different $(J,T,M_T)$ blocks. 
Calculations were performed in a $e_\text{max}=4$ single-particle space. 
Different data sets correspond to the natural grouping (yellow color grading) and to the '1432' grouping (blue color grading) of indices.}
\label{fig:svd}
\end{figure}

\begin{figure}[t!]
\includegraphics{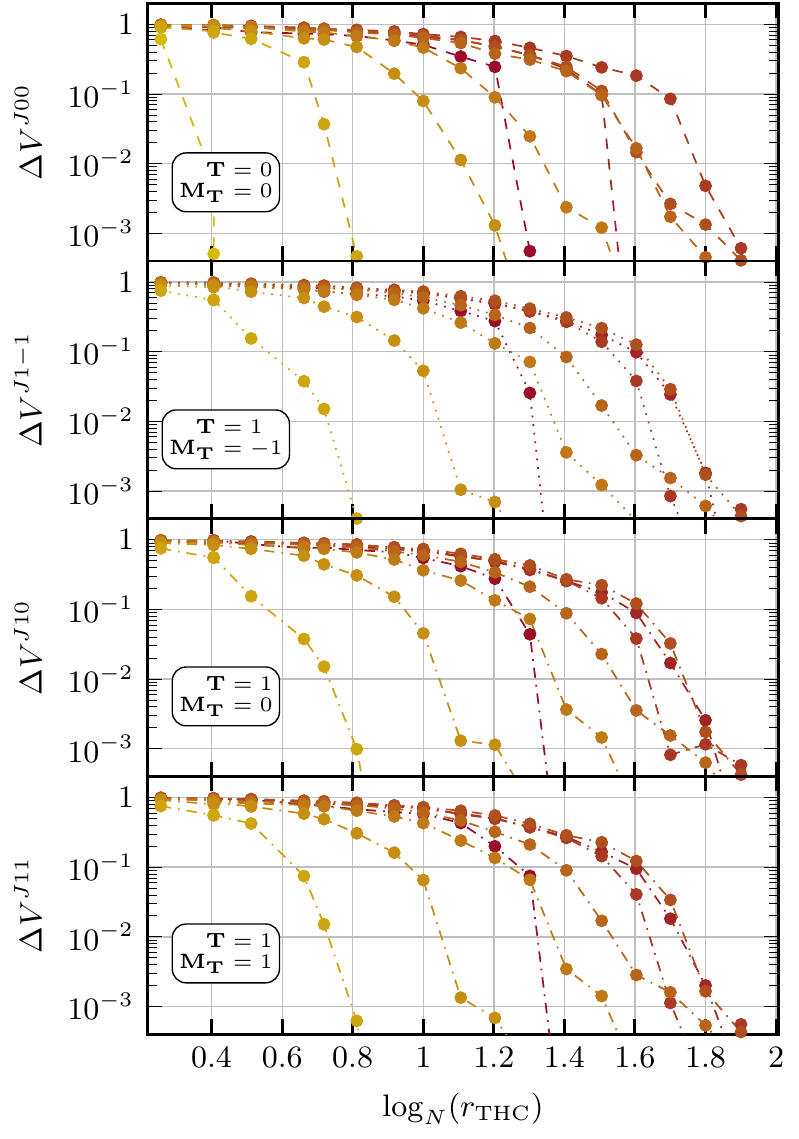}
\caption{(Color online) Same as Fig.~\ref{fig:cpd} for the THC format.}
\label{fig:thc}
\end{figure}

\subsubsection{Tensor hypercontraction}

The THC format is now employed to decompose the 2N interaction tensor. Since THC corresponds to a more sophisticated decomposition, lower ranks are expected to reach the same accuracy as with CPD. The numerical specifications to perform CPD on the left and right components defined in~\eqref{eq:cpdthc} are the same as for the bare CPD on the full interaction tensor.

The first step consists of investigating the SVD achieved in each $(J,T,M_T)$ block. In this respect, there exists a {\it natural} grouping of the first two and the last two indices together, respectively, i.e. of the two-body bra and ket indices in the $JT$-coupled matrix elements. However, another grouping is presently considered by pairing the first and fourth indices together as well as the third and second indices together. It is denoted as the \emph{'1432'} grouping. 

Figure~\ref{fig:svd} displays singular values on a semi-logarithmic plot in decreasing size. The four panels correspond to the four possible $(T,M_T)$ values. In each panel, a curve is associated to given $J$ value. One network of curves relates to the natural grouping whereas another one relates to the '1432' grouping.

Starting with the natural grouping, singular values decrease exponentially up to a maximum index $K_\text{max}$ that depends on $J$. For $K> K_\text{max}$, singular values drop to zero up to numerical accuracy. The value of $K_\text{max}$ is again linked to the number of non-zero elements of the interaction tensor in a given $(J,T,M_T)$ block. For $J=4$, the largest values of $K_\text{max}$ are of the order of one third of the matrix size whereas for other $J$ values it drops to one tenth of the matrix size. Truncating singular values with $K > K_\text{max}$, i.e., setting $r_\text{SVD} =K_\text{max}$, yields a very efficient way of pre-processing the interaction tensor and compressing the left and right singular matrices used as input for the CPD step. It indeed minimizes the range of the third index in Eq.~\ref{mode3CPD} and thus reduces the cost of the CPD step very significantly. Because $r_\text{SVD}$ strongly correlates with the sparsity of the initial tensor, one anticipates the pre-processing via the truncated SVD to be even more efficient for larger values of $e_\text{max}$ employed in realistic ab initio calculations.

It was observed in quantum chemistry that the '1432' grouping of indices yields singular values that do not decay exponentially as  a result of the long-range correlations in the electron repulsion tensor. Analogous results hold for the 2N interaction tensor as is visible from Fig.~\ref{fig:svd}. Singular values decrease only very slowly such that an efficient truncation is not possible. 

Finally, Fig.~\ref{fig:thc} provides the relative decomposition error as a function of the THC rank in each block.
The most important observation is that THC provides the same accuracy as CPD for $r_\text{THC} \ll r_\text{CPD}$. For $r_\text{THC} = N^{1.8}$ the relative error drops below $\epsilon\approx10^{-2}$ even in the worst-case scenario.
Again the THC rank for dense blocks is higher than for sparse ones, e.g., for the highest $J$ values an accuracy of $\epsilon=10^{-2}$ is obtained for $r_\text{THC}=N^{1.3}$. In all cases the THC ranks corresponding to $\epsilon=10^{-2}$ are below the critical value  of $N^2$, thus, yielding proper data compression. Also the THC ranks are independent of $M_T$ indicating again  that the key variable is the sparsity of the tensor and not the particular value of its individual entries.

The precision of the tensor decomposition and, in particular, the scaling of the THC rank as a function of the one-body basis size differs from observations in quantum chemistry where, in the worst-case scenarios, $r_\text{THC} \approx N^{1.6}$ yields an accuracy of $\epsilon=10^{-4}$~\cite{Schu17}. This difference is not surprising given that J-coupled matrix elements have many less zero entries compared to their $m$-scheme counterparts. Therefore, a clean comparison would require to perform the THC on the $m$-scheme nuclear interaction tensor. Since for $e_\text{max}=4$ the $m$-scheme dimension is already $\dim(\mathcal{H}_1) =140$, it is computationally very challenging at the moment.

Eventually, the THC format is considered to be superior to the CPD format for decomposing the nuclear interaction tensor. Therefore, the subsequent analysis  is restricted to the THC decomposition.

\subsection{$J$-coupled matrix elements in the HF basis}
\label{2bodyJcoupledHF}

\begin{figure}[t!]
\includegraphics[width=1.0\columnwidth]{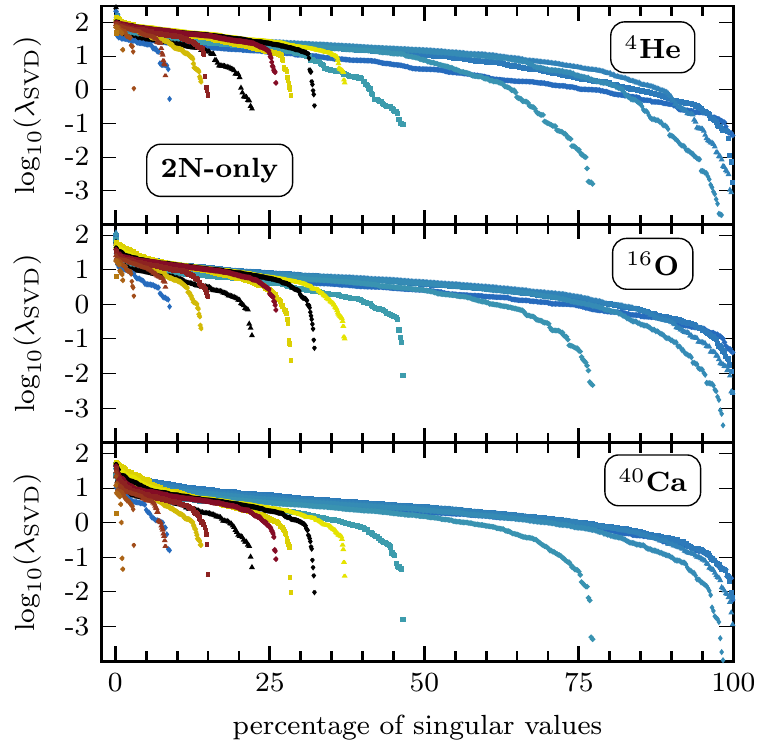}
\caption{(Color online) Size of the singular values of $H^J$ with 2N interactions only expressed in the HF basis for $^4$He, $^{16}$O and $^{40}$Ca. Different colors correspond to different $J$ values. Calculations are performed in an $e_\text{max}=4$ single-particle space. 
Different data sets correspond to the natural grouping (yellow color grading) and the '1432' grouping (blue color grading) of indices.}
\label{fig:svd_HF}
\end{figure}

\begin{figure}[t!]
\includegraphics[width=1.0\columnwidth]{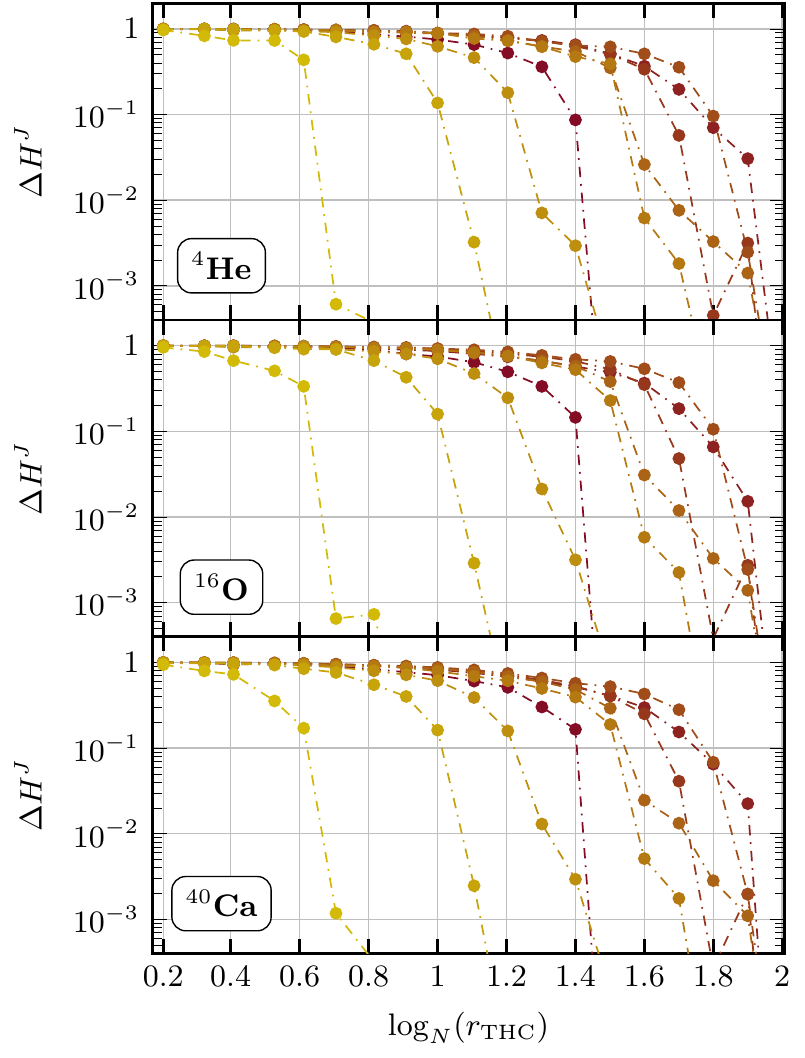}
\caption{(Color online) Relative error of the THC decomposition of $H$ with 2N interactions only expressed in the HF basis for $^4$He, $^{16}$O and $^{40}$Ca. Calculations are performed in an $e_\text{max}=4$ single-particle space. The same color convention as before is used for the different $J$ values.}
\label{fig:thc_J}
\end{figure}

In most many-body methods the one-body basis is conveniently obtained from a prior optimization of a mean-field reference state, e.g., by performing a Hartree-Fock calculation\footnote{For single-reference MBPT applications in closed-shell systems the use of a HF reference state is crucial to obtain a converging perturbative series~\cite{Ti16,Hu16}.}. In this section the tensor decomposition of the intrinsic nuclear Hamiltonian restricted to 2N forces
\begin{align}
H = T_\text{int}+ V \, ,
\end{align}
is discussed. The HO matrix elements of $H$ are transformed according to
\begin{align}
H^{\text{HF}}_{k_1k_2 k_3 k_4} = \sum_{l_1 l_2 l_3 l_4} C_{k_1 l_1} C_{k_2 l_2} C_{k_3 l_3} C_{k_4 l_4} H^{\text{HO}}_{l_1l_2 l_3 l_4} \, ,
\label{eq:hftrafo}
\end{align}
where $C_{kl}$ denotes the HF expansion coefficients, i.e., components of the eigenvectors of the Fock matrix. This transformation is system dependent such that the tensor decomposition of HF matrix elements has to be done for each nucleus under consideration\footnote{In App.~\ref{sec:hf}, the alternative scheme in which the Hamiltonian tensor is first decomposed in the HO basis and then transformed to the HF basis is briefly discussed. Given that this is only useful if 3N interactions can be treated in full (see Sec.~\ref{sec:no2b} for details of the current treatment), we do not proceed in this way and simply provide the corresponding discussion as a reference for the future.}. In order to ensure the robustness of our conclusions, the analysis is presently performed for three closed-shell nuclei covering the regions of light- and medium-mass systems, i.e. $^{4}$He, $^{16}$O and $^{40}$Ca.

\begin{figure}[t!]
\includegraphics[width=1.0\columnwidth]{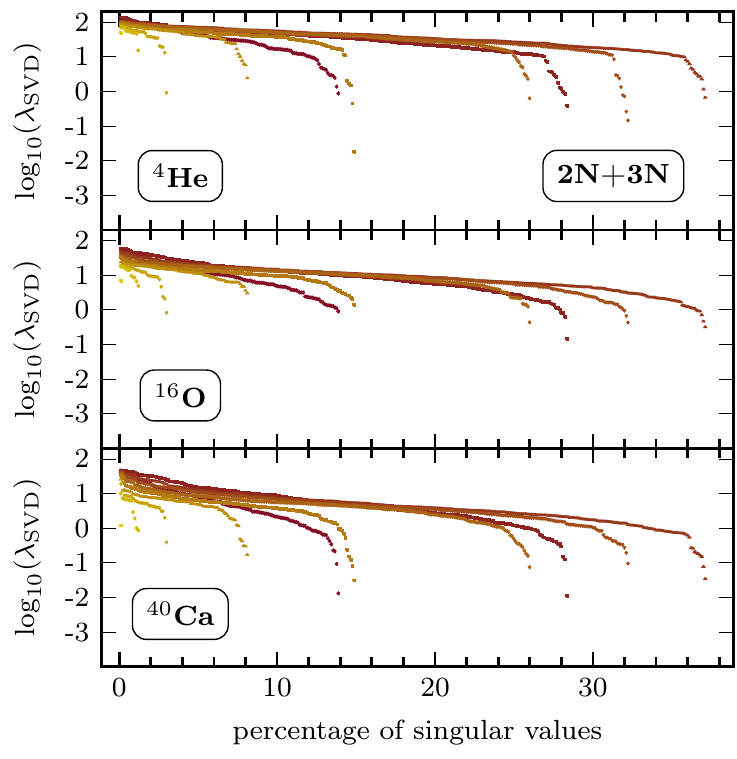}
\caption{(Color online) Same as Fig.~\ref{fig:svd_HF} for the nuclear Hamiltonian containing the contribution from the 3N interaction in the NO2B approximation.
Note that only the results corresponding to the natural grouping of indices are displayed with a modified horizontal scale.
}
\label{fig:svd_HF_3B}
\end{figure}
\begin{figure}[t!]
\includegraphics[width=1.0\columnwidth]{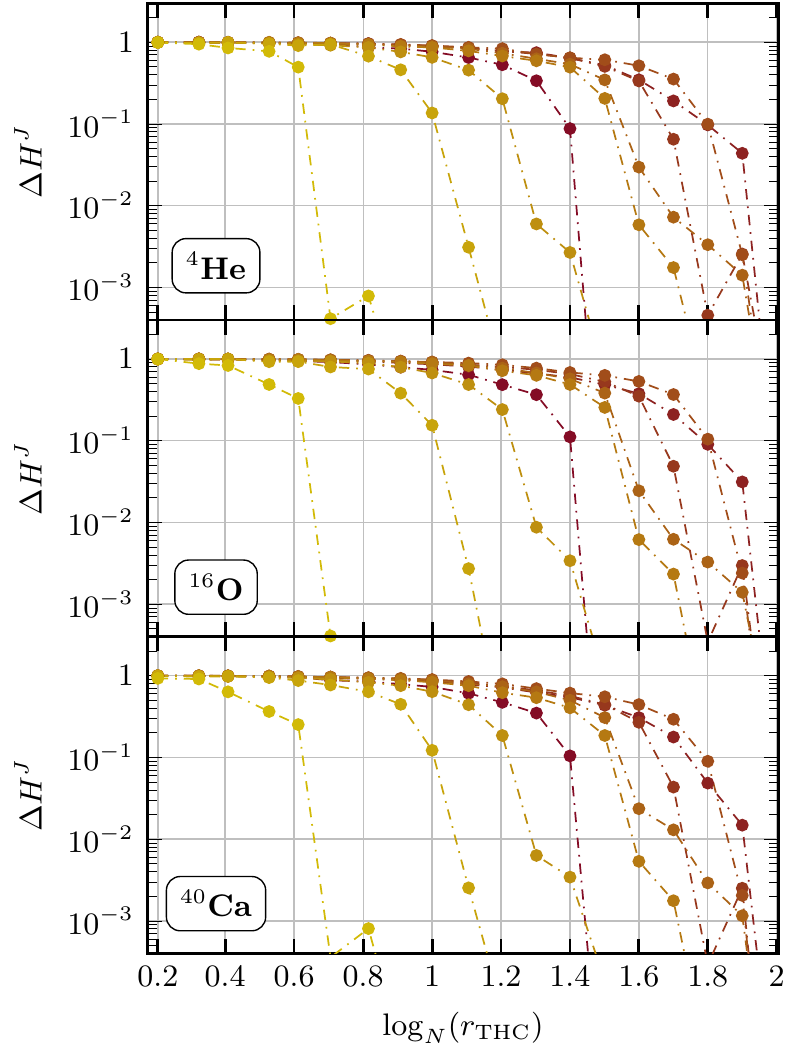}
\caption{(Color online) Same as Fig.~\ref{fig:thc_J} for the nuclear Hamiltonian containing the contribution from the 3N interaction in the NO2B approximation. Note that only the results corresponding to the natural grouping of indices are displayed with a modified horizontal scale.}
\label{fig:thc_J_3B}
\end{figure}

The resulting HF matrix elements, however, cannot be stored in $JT$-coupled scheme but require a decoupled form with respect to isospin quantum numbers, i.e., the use of a bare $J$-coupling scheme
\begin{align}
H^{J}_{\tilde \imath_1 \tilde \imath_2 \tilde \imath_3 \tilde \imath_4} \equiv \la \tilde \imath_1 \tilde \imath_2 (J) | H | \tilde \imath_3 \tilde  \imath_4 (J) \ra\, ,
\label{eq:inttensJ}
\end{align}
where indices denote now the collective set of quantum numbers
\begin{align}
\tilde \imath \equiv (n_i, l_i, j_i, t_i) \, ,
\end{align}
where the isospin projection $t_i$ distinguishes protons and neutrons. For a given single-particle basis dimension, the index range within a $J$ block $H^J$ is doubled compared to $V^{JTM_T}$. At the same time, the number of tensors to consider is reduced by a factor of four since the different $(T,M_T)$ channels are not distinguished.

As for the HO matrix elements, the efficiency of the truncated SVD is discussed first. Figure~\ref{fig:svd_HF} displays singular values for different nuclei and $J$ values both with respect to the natural grouping and the '1432' grouping. The same behaviour is observed as for $JT$-coupled HO matrix elements of the two-body interaction. For the natural grouping, singular values can be compressed down to $30\%$ for intermediate $J$ values and to much less than $10\%$ for extremal $J$. Contrarily, the '1432' grouping does not allow for any efficient compression of the tensor. One note that the results are qualitatively and quantitatively similar for $^{4}$He, $^{16}$O and $^{40}$Ca. 

Figure~\ref{fig:thc_J} provides an overview of the THC decomposition for the same benchmark systems. For high $J$ values, the required THC rank necessary to obtain a given accuracy is lower. For intermediate values $1\leq J\leq 5$ the THC ranks are rather high and the relative error is between $\Delta H^{J}=10^{-1}-10^{-2}$ for $r_\text{THC}\approx N^{1.8}$.

\subsection{Normal-ordered three-nucleon interaction}
\label{sec:no2b} 

It is a well-known feature that the restriction to 2N interactions fails to reproduce basic aspects of the phenomenology of nuclei and of extended nuclear matter. By including 3N interactions this problem can be qualitatively overcome such that a quantitative description can be achieved in light nuclei and envisioned for heavier systems.
However, in most many-body approaches the handling of full three-body operators is computationally too demanding at the moment such that one resorts to the so-called \emph{normal-ordered two-body approximation} (NO2B)
\begin{align}
\Theta_{pqrs} = \sum_{kl} w_{pqkrsl} \rho_{kl} \, ,
\label{eq:no2b}
\end{align}
yielding an effective two-body interaction (i.e. a mode-4 tensor) via an in-medium contraction of the three-body matrix elements $w_{pqrsut}$ (i.e. a mode-6 tensor) with the one-body density $\rho_{kl}$ of the quantum state under consideration.
In this way, dominant effects of 3N interactions are included without explicitly resorting to three-body operators. In no-core shell model (NCSM) calculations the induced error was estimated to be $1-3 \%$ in medium-light systems~\cite{RoBi12,Geb16}.

Working in the NO2B approximation puts us in position to apply tensor decomposition techniques in presence of 3N interactions  without having to explicitly decompose a mode-6 tensor. We thus repeat the analysis presented in Sec.~\ref{2bodyJcoupledHF} in this context. The one-body density matrix entering Eq.~\eqref{eq:no2b} is obtained from a prior HF optimization performed in the presence of the full 3N interaction. As visible from Fig.~\ref{fig:svd_HF_3B},  singular values behave as in the 2N interaction case, such that an efficient truncation can be achieved for the natural grouping. While not shown in Fig.~\ref{fig:svd_HF_3B}, we indicate that the '1432' grouping offers again no significant compression.

For the THC decomposition, the same pattern as for 2N interactions only is observed once again. In particular, Figs.~\ref{fig:thc_J} and~\ref{fig:thc_J_3B} are essentially identical. A relative error of $\Delta H^{J} = 10^{-1}-10^{-2}$ is achieved for $r_\text{THC}=N^{1.8}$ while for sparse $J$ blocks taking $r_\text{THC} \leq N^{1.5}$ is typically sufficient.
This ends to demonstrate that a systematically improvable decomposition of state-of-the-art nuclear Hamiltonians can be performed in presence of 3N forces.

\subsection{Overview}
\label{sec:overview} 

\begin{table}
\begin{tabular}{c | c | c | c | c | c | c}
basis & \, tensor \, & \, format \, & \, coupling\,  & \, N \, & \, $r_{\text{min}}$ \, & \, $r_{\text{max}}$ \,  \\
\hline \hline
HO & $V^{JTM_T}$ & CPD & $JT$-coupled & 15 & $N^{1.6}$ & $N^{2.4}$ \\
HO & $V^{JTM_T}$ & THC & $JT$-coupled & 15 & $N^{0.4}$ & $N^{1.7}$ \\
HF & $H^{J}$ & THC  & $J$-coupled & 30 & $N^{0.7}$ & $N^{1.8}$ \\
\hline \hline
\end{tabular}
\label{tab:overview}
\caption{Results obtained from different tensor formats and single-particle bases along with different angular-momentum and isospin coupling schemes. The quantities $r_\text{min}$ and $r_\text{max}$ correspond to the minimal and maximal THC decomposition ranks selected for different $J$ or $(J,T,M_T)$ channels whenever fixing the decomposition error  $\Delta V^{JTM_T}$ or $\Delta H^{J}$ to be $\epsilon=0.1$. All results refer to a model space associated with $e_\text{max}=4$ and include 2N interactions only (including the 3N force in the NO2B approximation does not modify the results associated with the last row).}
\end{table}

An overview of the results is provided in Tab.~\ref{tab:overview}. For a decomposition error $\Delta V^{JTM_T}$ or $\Delta H^{J}$ set to $\epsilon=10^{-1}$, the minimal and maximal ranks obtained among all $J$ values are listed. The superiority of THC over CPD in this respect is clearly visible. Including or not the 3N interaction via the NO2B approximation  does not modify the results associated with the last row.

\subsection{Data compression}
\label{sec:datacomp} 

\begin{figure*}
\centering
\scalebox{1.1}{
\includegraphics{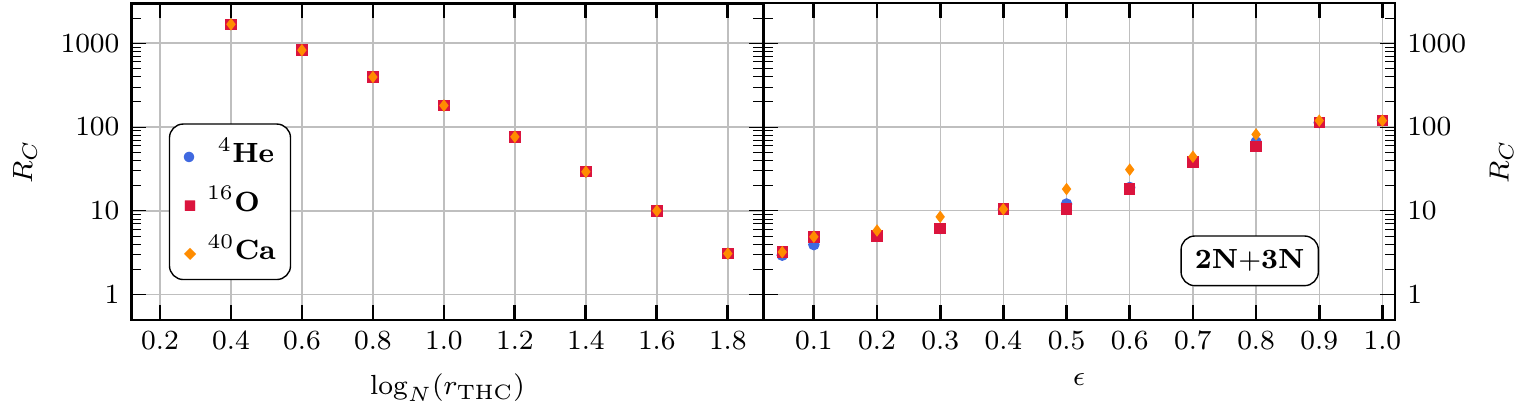}
}
\label{fig:datacomp}
\caption{(Color online) Data compression factor $R_C$ reached by the THC decomposition of the nuclear Hamiltonian. Calculations are performed in a $e_\text{max}=4$ single-particle space with a chiral Hamiltonian expressed in the HF basis and including the 3N interaction in the NO2B approximation. Left panel: the THC decomposition is performed using the same rank in all $J$ blocks. Right channel: the THC decomposition is performed using the same error in all $J$ blocks.
}
\end{figure*}

To illustrate the memory gain obtained via the THC, a data compression factor $R_C$ is now displayed in Fig.~\ref{fig:datacomp} for the Hamiltonian expressed in the HF basis and including the 3N interaction via the NO2B approximation. The  factor $R_C$ is defined as the ratio of the number of entries needed to store the initial interaction or Hamiltonian tensor over the number needed to store its factorized version.  Given that the THC rank depends on the $J$ value of a given block, the approximation on the overall tensor is set in two different ways. The right panel of Fig.~\ref{fig:datacomp} displays $R_C$ as a function of an error threshold $\epsilon$ common to all $J$ channels. This corresponds to using $J$-dependent THC ranks obtained by setting $\Delta H^J = \epsilon$. The left panel of Fig.~\ref{fig:datacomp} displays $R_C$ as a function of a THC rank $r_\text{THC}$ common to all $J$ blocks. This corresponds to using different errors $\Delta H^J$ in each $J$ block. 

One first observes that the compression factor decreases with the increasing accuracy in both cases, i.e. for increasing $r_\text{THC}$ or decreasing $\epsilon$. Of course, $R_C=1$ in the limit where no approximation is made on the Hamiltonian tensor.  Next, one notices that the trend is monotonous, independently of the way the approximation on the overall tensor is set. On the left panel, it appears clearly that the compression factor is nucleus independent. Indeed, setting a common rank for all $J$ blocks fixes by construction the number of entries in the factorized form of the tensor, which is system independent as long as one uses the same $e_\text{max}$ as is presently done. Given that the error $\epsilon$ actually probes the values of the tensor entries, and thus impacts the rank of the associated factorized tensor in each $J$ block, the curves on the right panel are not identical. Still, the nucleus dependence is essentially non existent.

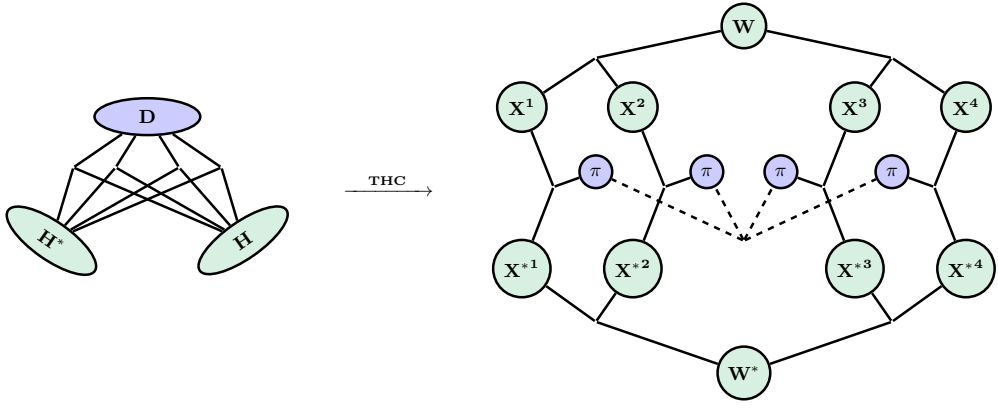
\begin{figure*}[t!]
\scalebox{.8}{
\begin{tikzpicture}[thick,scale=0.5]
\newcommand{\lwidth}{1.2pt}
\newcommand{\thcdh}{35pt}
\newcommand{\thcdv}{15pt}
\newcommand{\lusep}{50pt}
\newcommand{\xden}{35pt}
\node[circle,draw,line width=\lwidth,fill=mediumseagreen!20] (w) at (current page.center) {$\mathbf{W}$};
\node[circle,draw,line width=\lwidth,fill=mediumseagreen!20,below =1*\thcdv  of w,xshift=-3.0*\thcdh] (y1) {$\mathbf{X^1}$};
\node[circle,draw,line width=\lwidth,fill=mediumseagreen!20,below =1*\thcdv  of w,xshift=-1.5*\thcdh] (y2) {$\mathbf{X^2}$};
\node[circle,draw,line width=\lwidth,fill=mediumseagreen!20,below =1*\thcdv  of w,xshift=1.5*\thcdh] (y3) {$\mathbf{X^3}$};
\node[circle,draw,line width=\lwidth,fill=mediumseagreen!20,below =1*\thcdv  of w,xshift=3.0*\thcdh] (y4) {$\mathbf{X^4}$};
\node[inner sep=0pt, outer sep=0pt, above=0.7*\thcdv of y2, xshift= -0.5*\thcdh] (cpd1) {};
\node[inner sep=0pt, outer sep=0pt, above =0.7*\thcdv of y3, xshift= 0.5*\thcdh] (cpd2) {};
\draw[line width=\lwidth] (cpd1) -- (y1);
\draw[line width=\lwidth] (cpd1) -- (y2);
\draw[line width=\lwidth] (cpd1) -- (w);
\draw[line width=\lwidth] (cpd2) -- (y3);
\draw[line width=\lwidth] (cpd2) -- (y4);
\draw[line width=\lwidth] (cpd2) -- (w);
\node[circle,draw,line width=\lwidth,fill=mediumseagreen!20,below =\lusep of y1] (y1c) {$\mathbf{{X^*}^1}$};
\node[circle,draw,line width=\lwidth,fill=mediumseagreen!20,below =\lusep of y2] (y2c) {$\mathbf{{X^*}^2}$};
\node[circle,draw,line width=\lwidth,fill=mediumseagreen!20,below =\lusep of y3] (y3c) {$\mathbf{{X^*}^3}$};
\node[circle,draw,line width=\lwidth,fill=mediumseagreen!20,below =\lusep of y4] (y4c) {$\mathbf{{X^*}^4}$};
\node[circle,draw,line width=\lwidth,fill=mediumseagreen!20,below =1*\lusep+6*\thcdv of w] (wc) {$\mathbf{W^*}$};
\node[inner sep=0pt, outer sep=0pt, below=0.7*\thcdv of y2c,xshift=-0.5*\thcdh] (cpd3) {};
\node[inner sep=0pt, outer sep=0pt, below=0.7*\thcdv of y3c,xshift=0.5*\thcdh] (cpd4) {};
\draw[line width=\lwidth] (cpd3) -- (y1c);
\draw[line width=\lwidth] (cpd3) -- (y2c);
\draw[line width=\lwidth] (cpd3) -- (wc);
\draw[line width=\lwidth] (cpd4) -- (y3c);
\draw[line width=\lwidth] (cpd4) -- (y4c);
\draw[line width=\lwidth] (cpd4) -- (wc);
\node[circle,draw,line width=\lwidth,fill=blue!20,below =0.2*\lusep of y1,xshift=\xden] (pi1) {$\mathbf{\pi}$};
\node[circle,draw,line width=\lwidth,fill=blue!20,below =0.2*\lusep of y2,xshift=\xden] (pi2) {$\mathbf{\pi}$};
\node[circle,draw,line width=\lwidth,fill=blue!20,below =0.2*\lusep of y3,xshift=-\xden] (pi3) {$\mathbf{\pi}$};
\node[circle,draw,line width=\lwidth,fill=blue!20,below =0.2*\lusep of y4,xshift=-\xden] (pi4) {$\mathbf{\pi}$};
\node[inner sep=0pt, outer sep=0pt,below =0.5*\lusep of y1,xshift=15pt] (i) {};
\node[inner sep=0pt, outer sep=0pt,below =0.5*\lusep of y2,xshift=15pt] (j) {};
\node[inner sep=0pt, outer sep=0pt,below =0.5*\lusep of y3,xshift=-15pt] (a) {};
\node[inner sep=0pt, outer sep=0pt,below =0.5*\lusep of y4,xshift=-15pt] (b) {};
\draw[line width=\lwidth] (y1) -- (i);
\draw[line width=\lwidth] (y1c) -- (i);
\draw[line width=\lwidth] (pi1) -- (i);
\draw[line width=\lwidth] (y2) -- (j);
\draw[line width=\lwidth] (y2c) -- (j);
\draw[line width=\lwidth] (pi2) -- (j);
\draw[line width=\lwidth] (y3) -- (a);
\draw[line width=\lwidth] (y3c) -- (a);
\draw[line width=\lwidth] (pi3) -- (a);
\draw[line width=\lwidth] (y4) -- (b);
\draw[line width=\lwidth] (y4c) -- (b);
\draw[line width=\lwidth] (pi4) -- (b);
\node[inner sep=0pt,outer sep=0pt,below =6*\thcdv of w] (laplace) {};
\draw[dashed,line width=\lwidth] (laplace) -- (pi1);
\draw[dashed,line width=\lwidth] (laplace) -- (pi2);
\draw[dashed,line width=\lwidth] (laplace) -- (pi3);
\draw[dashed,line width=\lwidth] (laplace) -- (pi4);
\node[ellipse,draw,line width=\lwidth,minimum width =50pt,fill=blue!20,below left =1cm and 9cm of w] (D) {$\mathbf{D}$};
\node[rotate=-35,ellipse,draw,line width=\lwidth,minimum width =50pt,fill=mediumseagreen!20,below left =2cm and 0.3cm of D] (Hc) {$\mathbf{H^*}$};
\node[rotate=35,ellipse,draw,line width=\lwidth,minimum width =50pt,fill=mediumseagreen!20,below right =2cm and 0.3cm of D] (H) {$\mathbf{H}$};
\node[inner sep=0pt, outer sep=0pt,below =0.5cm of D,xshift=-35pt] (a) {};
\node[inner sep=0pt, outer sep=0pt,below =0.5cm of D,xshift=-15pt] (b) {};
\node[inner sep=0pt, outer sep=0pt,below =0.5cm of D,xshift=15pt] (c) {};
\node[inner sep=0pt, outer sep=0pt,below =0.5cm of D,xshift=35pt] (d) {};
\draw[line width=\lwidth] (H) -- (a);
\draw[line width=\lwidth] (Hc) -- (a);
\draw[line width=\lwidth] (D) -- (a);
\draw[line width=\lwidth] (H) -- (b);
\draw[line width=\lwidth] (Hc) -- (b);
\draw[line width=\lwidth] (D) -- (b);
\draw[line width=\lwidth] (H) -- (c);
\draw[line width=\lwidth] (Hc) -- (c);
\draw[line width=\lwidth] (D) -- (c);
\draw[line width=\lwidth] (H) -- (d);
\draw[line width=\lwidth] (Hc) -- (d);
\draw[line width=\lwidth] (D) -- (d);
\node[below right =0.6cm and 2.5cm of D] (arrow) {$\mathbf{\xrightarrow{\quad \textbf{THC} \quad}}$};
\end{tikzpicture}
}
\label{fig:mp2tn}
\caption{Tensor network of the second-order energy equation in its original (left) and factorized (right) forms. Objects in green correspond to the THC factors of the intrinsic Hamiltonian whereas objects in blue correspond to the CPD factors generated from the inverse Laplace transform of the energy denominators. }
\end{figure*}

\section{Ground-state energies}
\label{sec:gse}

To benchmark the performance of tensor-decomposed tensors in many-body calculations, second-order MBPT calculations of closed-shell nuclei are performed as a simple testbed. 

\subsection{Second-order many-body perturbation theory}

The central quantity of present interest is the second-order (Rayleigh-Schr\"odinger) MBPT correction to the ground-state energy  obtained from the so-called M{\o}ller-Plesset partitioning~\cite{SzOs82,ShBa09}
\begin{align}
E^{(2)} = -\frac{1}{4} \sum_{abij} \frac{ H_{abij} H_{ijab}}{\epsilon_a + \epsilon_b - \epsilon_i - \epsilon_j} \, ,
\label{eq:mbpt2}
\end{align}
where roman labels $a,b$ and $i,j$ denote particle and hole states, i.e., single-particle states that are unoccupied and occupied in the HF reference Slater determinant, respectively. Furthermore, $\epsilon_k$ refers to HF single-particle energies. Equation~\eqref{eq:mbpt2} is the leading correction providing the bulk part of dynamic correlation effects in closed-shell nuclei when starting from SRG-evolved chiral Hamiltonians~\cite{Ti16,Hu17}.

Making use of angular-momentum coupling techniques, Eq.~\eqref{eq:mbpt2} is rewritten under the working form
\begin{align}
E^{(2)} = -\frac{1}{4} \sum_{J} \hat J^2 \sum_{\tilde a\tilde b\tilde \imath \tilde \jmath} \frac{ H^J_{\tilde a\tilde b\tilde \imath \tilde \jmath} H^J_{\tilde \imath \tilde \jmath\tilde a\tilde b}}{\epsilon_{\tilde a} + \epsilon_{\tilde b} - \epsilon_{\tilde \imath} - \epsilon_{\tilde  \jmath}} \, ,
\label{eq:mbpt2J}
\end{align}
where $\hat x \equiv \sqrt{2x+1}$~\cite{VaMo88}. Working with a spherically-restricted HF solution, single-particle energies are $m$-independent, i.e., $\epsilon_{\tilde p} = \epsilon_p$.

The computational complexity of a many-body framework is related to the number of internal summations required to evaluate the contractions between the tensors. Second-order HF-MBPT involves two particle and two hole summations, i.e.
\begin{align}
E^{(2)} \sim n_p^2 \, n_h^2\, ,
\end{align}
where $n_p$ and $n_h$ denotes the number of particle and hole states, respectively. Without distinguishing them, second-order MBPT is thus loosely referred to as a $N^4$ process. More advanced state-of-the-art many-body approaches typically involve a $N^6$ scaling, e.g., third-order MBPT, CC truncated at the singles and doubles level (CCSD), IM-SRG truncated at the two-body level (IMSRG(2)) or Green's function at the ADC(3) level. For high-accuracy calculations a final, e.g. $N^7$, evaluation is performed to approximately account for higher-order effects in a non-iterative way\footnote{The present analysis explicitly excludes valence-space approaches requiring the diagonalization of a dressed Hamiltonian in a truncated valence space. In this case the exponential scaling of the final diagonalization convolutes with the building of the valence-space Hamiltonian that is performed at (low) polynomial cost.}.

\subsection{Many-body tensor factorization}

As alluded to in the introduction, many-body calculations typically involve the contraction of the Hamiltonian tensor with what can be denoted as {\it many-body tensors} whose detailed form is characteristic of the many-body formalism of interest. In order to exploit the factorized form of the Hamiltonian tensor, the involved n-tuple sum must be disentangled. In order to achieve this goal, the many-body formalism must be reformulated in terms of factorized many-body tensors. 

In this respect, MBPT(2) provides the simplest and yet illustrative example. As Eqs.~\eqref{eq:mbpt2} and~\eqref{eq:mbpt2J} make clear, computing $E^{(2)}$ involves the contraction of (twice) the Hamiltonian tensor with
\begin{align}
D^{ab}_{ij} \equiv \frac{1}{\epsilon_a + \epsilon_b - \epsilon_i - \epsilon_j} \, .
\end{align}
A decomposition of this mode-4 tensor is easily obtained due to its known analytical structure. The situation is less trivial in non-perturbative methods where the many-body tensors at play constitute unknowns to be obtained along the way. In the present case, $D$ can rewritten exactly via the inverse Laplace transform according to
\begin{align}
\frac{1}{\epsilon_a + \epsilon_b - \epsilon_i - \epsilon_j} = \int_0^\infty e^{-t(\epsilon_a + \epsilon_b - \epsilon_i - \epsilon_j)} dt \, ,
\end{align}
which leads, via a numerical quadrature scheme, to
\begin{align}
D^{ab}_{ij} & \approx \sum_{s=-M}^{+M} \omega_s e^{-t_s(\epsilon_a + \epsilon_b - \epsilon_i - \epsilon_j)} \nonumber \\
&\equiv \sum_{s=-M}^{+M}   \pi_{as} \, \pi_{bs} \, \omega_s \, \pi_{si} \, \pi_{sj} \, ,
\label{eq:dencpd}
\end{align}
where by convention $\pi_{sp} = (\pi_{ps})^{-1}$. The rank of Eq.~\eqref{eq:dencpd} is $r_D\equiv 2M+1$.
Numerical values of $\omega_s$ and $t_s$ are tabulated in the literature~\cite{BrHa05} such that high precision can be reached with very few grid points. In particular the size of the integration mesh was shown to be independent of the system. Note that Eq.~\eqref{eq:dencpd} is equivalent to performing a CPD of  $D$. Since in closed-shell systems the HF reference Slater determinant displays a pronounced shell gap at the Fermi energy, the entries of $D$ are all non-vanishing and, therefore, the quadrature is well-defined.

\subsection{MBPT(2) tensor network}

With the decomposed tensors at hand, the tensor network associated with $E^{(2)}$ can be formulated. It is displayed graphically in Fig.~\ref{fig:mp2tn} in its original and in its tensor-factorized forms. Although every $\{X^{i},W\}$ factor carries an additional angular-momentum label $J$, the corresponding labels are left out in the figure since the different $J$ blocks do not mix in Eq.~\eqref{eq:mbpt2J}.

The second-order energy correction is now written as
\begin{align}
E^{(2)} &= 
-\frac{1}{4} \sum_{J \alpha \beta  \gamma \delta s} \hspace{-4pt}\hat J^2 
{^{J\hspace{-2pt}}  W}_{\alpha \beta} 
{^{J\hspace{-2pt}}  W}_{\gamma \delta}
{^{J\hspace{-2pt}} A}^{\alpha \gamma}_s
{^{J\hspace{-2pt}} B}^{\alpha \gamma}_s
{^{J\hspace{-2pt}} C}^{\beta \delta}_s
{^{J\hspace{-2pt}} D}^{\beta \delta}_s \, , \nonumber
\end{align}
where the intermediates
\begin{subequations}
\allowdisplaybreaks
\begin{align}
{^{J\hspace{-2pt}} A}^{\alpha \gamma}_s &\equiv \sum_a {^J X}^1_{a \alpha} {^J X}^1_{a \gamma} \pi_{a s} \,, \\
{^{J\hspace{-2pt}} B}^{\alpha \gamma}_s &\equiv \sum_b {^J X}^2_{b \alpha} {^J X}^2_{b \gamma} \pi_{b s} \, ,\\
{^{J\hspace{-2pt}} C}^{\beta \delta}_s &\equiv \sum_i {^J X}^3_{i \beta} {^J X}^3_{i \delta} \pi_{si} \, , \\
{^{J\hspace{-2pt}} D}^{\beta \delta}_s &\equiv \sum_j {^J X}^4_{j \beta} {^J X}^4_{j \delta} \pi_{sj} \, ,
\end{align}
\end{subequations}
have been introduced. The evaluation costs of ($A,B$) and ($C,D$) are  $\mathcal{O} (r^2_\text{THC} \cdot n_h \cdot r_D)$ and $\mathcal{O} (r^2_\text{THC} \cdot n_p \cdot r_D)$, respectively. Defining new intermediates via
\begin{subequations}
\begin{align}
\Jtens{M}{J}{\alpha \delta}{s} &\equiv \sum_{\gamma}\Jtens{A}{J}{\alpha \gamma}{s}  \Jtens{B}{J}{\alpha \gamma}{s} {^{J\hspace{-2pt}}  W}_{\gamma \delta} \, , \\
\Jtens{N}{J}{\alpha \delta}{s} &\equiv \sum_{\beta} \Jtens{C}{J}{\beta \delta}{s}  \Jtens{D}{J}{\beta \delta}{s} {^{J\hspace{-2pt}}  W}_{\alpha \beta}  \, ,
\end{align}
\label{eq:r3int}
\end{subequations}
one is eventually left with
\begin{align}
E^{(2)} = -\frac{1}{4} \sum_{J} \hspace{-2pt} \hat J^2 \sum_s \sum_{\alpha \delta} \Jtens{M}{J}{\alpha \delta}{s} \Jtens{N}{J}{\alpha \delta}{s} \, . \label{finalMBPT2}
\end{align}

\begin{figure*}
\centering
\scalebox{1.1}{
\includegraphics{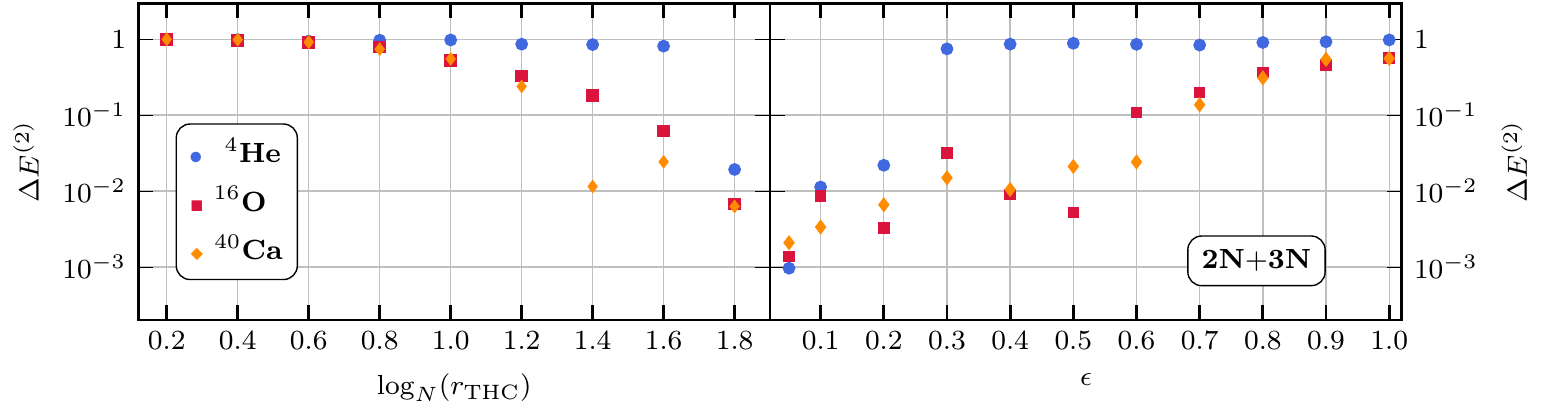}
}
\label{fig:mp2gs}
\caption{(Color online) Relative error $\Delta E^{(2)}$ of the tensor-decomposed second-order ground-state energy correction for $^4$He, $^{16}$O and $^{40}$Ca. Calculations are performed in a $e_\text{max}=4$ single-particle space with a chiral Hamiltonian expressed in the HF basis and including the 3N interaction in the NO2B approximation. Left panel: the THC decomposition is performed using the same rank in all $J$ blocks. Right channel: the THC decomposition is performed using the same error in all $J$ blocks.
}
\end{figure*}

While performing the contractions in Eq.~\ref{eq:r3int} is carried out in $\mathcal{O} (r^3_\text{THC} \cdot r_D)$, the summation over $(s,\alpha,\delta)$ in Eq.~\eqref{finalMBPT2} requires $\mathcal{O} (r^2_\text{THC} \cdot r_D)$ operations.  Obviously, the computational scaling of $E^{(2)}$ depends on the THC ranks $r_\text{THC}$ in each $J$ block, which themselves depend on the chosen approximation error $\epsilon$ on the Hamiltonian. As the most expensive contraction in Eq.~\eqref{eq:r3int} is a cubic polynomial in $r_\text{THC}$, $E^{(2)}_\text{THC}$ scales as
\begin{align}
E^{(2)}_\text{THC} \sim \mathcal{O} (N^{5.4})
\end{align}
for an accuracy of $\epsilon = 10^{-1}$ according to Tab.~\ref{tab:overview}. This is worse than the naive $\mathcal{O}(N^4)$ scaling of the original second-order correction. However, the aim of the present work is not to derive a low-scaling approximation of an already low-cost many-body method but rather to benchmark the propagation of the decomposition error of many-body tensors on nuclear observables (see Sec.~\ref{sec:mp2res} below). One has to move to more expensive methods to begin with to generate a reduction of the numerical scaling. In this context, it was of example claimed that third-order MBPT can be carried out in $\mathcal{O}(N^4)$ instead of the usual $\mathcal{O}(N^6)$ when using a THC-decomposed electron repulsion tensor~\cite{Ho12a}\footnote{In the cited paper there is no discussion on the scaling of the THC rank as a function of single-particle basis size. We expect that the authors assume a linear dependence $r_\text{THC} \sim N$. Additionally, it is to be noted that a different approach was used to obtain the THC factors involving density-fitting techniques.}. This analysis is not presently carried out since the $J$-coupled third-order formulas require an angular-momentum recoupling of the particle-hole diagram that is incompatible with the standard coupling order of $J$-coupled matrix elements. This will require an analysis of the THC ansatz for Pandya-transformed~\cite{Su07}, i.e. particle-hole-recoupled, matrix elements.

\subsection{Results}
\label{sec:mp2res}

To measure the impact of the tensor factorization, the relative error on the second-order energy correction is introduced
\begin{align}
\Delta E^{(2)} \equiv \frac{\big\vert E^{(2)}_\text{THC} - E^{(2)}\big \vert}{\big \vert E^{(2)} \big \vert} \, ,
\end{align}
which goes to zero in the limit of an exact THC decomposition. Because of the highly-accurate decomposition of the analytically known tensor $D$ in MBPT(2), the error presently propagates entirely from the approximation made on the Hamiltonian tensor. Figure~\ref{fig:mp2gs} displays $\Delta E^{(2)}$ as a function of the THC decomposition error for doubly closed-shell nuclei $^4$He, $^{16}$O and $^{40}$Ca. The Hamiltonian tensor is the same as the one employed in Secs.~\ref{sec:no2b} and~\ref{sec:datacomp}.

The right panel of Fig.~\ref{fig:mp2gs} displays $\Delta E^{(2)}$ as a function of the error $\epsilon$ on the Hamiltonian tensor introduced in Sec.~\ref{sec:datacomp}. A global trend is visible such that lower values of $\epsilon$ yield lower $\Delta E^{(2)}$. However, this behaviour is non-monotonic since a better global approximation of the Hamiltonian might still lead to larger deviations on the level of individual tensor entries, which might eventually affect the value of $\Delta E^{(2)}$. One futher observes that $^4$He displays 100$\%$ error all the way down to $\epsilon = 0.3$ before dropping abruptely to catch up with the smoother curves obtained for $^{16}$O and $^{40}$Ca. This peculiar behavior relates to the anomalously low correlation energy per particle in this very light and tightly bound system~\cite{Ti16}. Because our focus is on mid- and heavy-mass nuclei, the particular behavior observed for $^4$He is ignored when drawing general conclusions below. Eventually, a THC approximation error of $\epsilon \approx 10^{-1}$ is sufficient to obtain $\Delta E^{(2)} \approx 10^{-2}$ for the three nuclei under consideration. Therefore, even though the matrix elements are only approximated to an accuracy of $10^{-1}$, the precision on the observable of interest is one order of magnitude better. Even a quite crude approximation on the matrix elements thus yields an accuracy that is good enough to perform precision studies. Further decreasing the THC error to $\epsilon \approx 5 \times 10^{-2}$, accuracies of up $\Delta E^{(2)}\approx 10^{-3}$ can be reached.

\begin{figure*}
\centering
\scalebox{1.1}{
\includegraphics{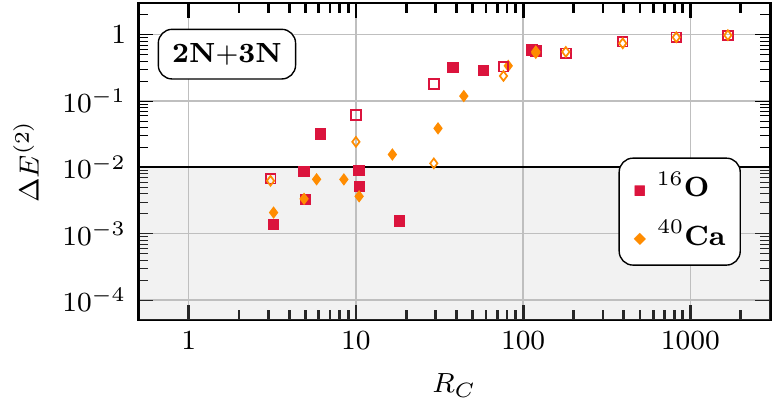}
}
\label{fig:correl}
\caption{(Color online) Relative error $\Delta E^{(2)}$ of the tensor-decomposed second-order ground-state energy correction against data compression factor $R_C$ for $^{16}$O and $^{40}$Ca. Calculations are performed in a $e_\text{max}=4$ single-particle space with a chiral Hamiltonian expressed in the HF basis and including the 3N interaction in the NO2B approximation. Full (empty) symbols are obtained by setting the THC approximation of the Hamiltonian tensor through $\epsilon$ ($r_\text{THC}$). The grey area indicates the region of sub-percent accuracy on the second-order ground-state correlation energy.
}
\end{figure*}

The left panel of Fig.~\ref{fig:mp2gs} displays $\Delta E^{(2)}$ as a function of the common rank $r_\text{THC}$ defined in Sec.~\ref{sec:datacomp}. While the general behavior is similar to the right panel, the trend is more monotonic. Eventually, a common THC rank $r_\text{THC} \approx N^{1.8}$ is sufficient to obtain $\Delta E^{(2)} \approx 10^{-2}$ for the three nuclei under consideration.

In order to combine the informations provided in Figs.~\ref{fig:datacomp} and~\ref{fig:mp2gs}, the correlation between $\Delta E^{(2)}$ and $R_C$ is displayed in Fig.~\ref{fig:correl} for both $^{16}$O and $^{40}$Ca. A clear correlation is observed, thus confirming the expectation that a compromise must be made between the needed accuracy and the data compression that can be reached. Furthermore, the correlation does not depend decisively on the variable used to set the approximation on the Hamiltonian tensor. Eventually, a sub-percent accuracy is typically reached while reducing the input data by one order of magnitude. While this number is presently obtained in a small single-particle model space, we expect the data compression achieved for a given accuracy on the observable to increase significantly with $e_\text{max}$.

\section{Discussion}

We conclude with a last set of comments
\begin{enumerate}
\item Applications presented in this work are restricted to angular-momentum-coupled matrix elements in a symmetry-adapted single-particle basis. Recently, it was shown that the use of a Bogoliubov reference state breaking $U(1)$ symmetry in Gorkov SCGF~\cite{SoCi13}, Bogoliubov CC~\cite{Si15} and Bogoliubov MBPT~\cite{Ti18b,Art18a} is highly beneficial to account for static correlation effects  in open-shell superfluid nuclei while still resorting to single-reference methods. While it is of strong interest to extend tensor factorizations to such frameworks it remains yet to be seen how the Hamiltonian matrix elements expressed in the symmetry-broken quasi-particle basis authorize efficient low-rank decompositions. In particular, while an efficient THC was presently shown to heavily rely on a \emph{natural grouping} of indices, no such natural grouping exists for matrix elements in the quasi-particle basis.  While the first tests in open-shell oxygen isotopes do not reveal such issues~\cite{TiDu18}, the analysis needs to be performed systematically. This work is under way. 
\item Appendix~\ref{sec:hf} briefly discusses that the THC decomposition of the HF matrix elements can be obtained by either performing the THC decomposition on the interaction tensor represented in the HF basis or by convoluting the THC factors of the interaction tensor generated in the HO basis with the HF expansion coefficients. Of course, the second option is highly preferable given that it is \emph{universal}, i.e., it relies on a system-independent factorization of the Hamiltonian that can be performed once for a chosen $e_\text{max}$ and from which nucleus-dependent quantities can be obtained via a basis transformation whenever necessary. However, the two steps of (i) decomposing the tensor and (ii) basis transformation are \emph{non-commutative} in principle given that the approximation on the Hamiltonian tensor can influence the results of the HF calculation. Therefore, it is not guaranteed that the same decomposition error is achieved in both cases. The comparison of both approaches needs to be investigated whenever the factorization of the mode-6 tensor associated to the 3N interaction becomes possible.
\item Results obtained in this paper reveals higher THC ranks for the nuclear Hamiltonian tensor than for the electron repulsion tensor in quantum chemistry. While not necessarily surprising given the different nature of the interactions under consideration, the difference regarding the efficiency of the tensor decomposition is believed to be mainly driven by the  coupling scheme employed. While the electron-electron repulsion tensor is stored in an '$m$-scheme analogue', the nuclear interaction is processed in $J$-coupled or $JT$-coupled schemes. This preprocessing of symmetry properties lowers the (relative) number of non-zero entries in the tensor, which eventually leaves less redundancies to exploit in the factorization. Correspondingly, it is of interest to benchmark the THC decomposition on $m$-scheme matrix elements in the near future. This will anyway be necessary to employ tensor-decomposition techniques in many-body methods allowing for the breaking (and the restoration) of $SU(2)$ symmetry in doubly open-shell nuclei in the future.
\item A computational bottleneck of the THC decomposition is the large range spanned by two of the indices of the left and right singular vectors used as an input to the CPD decomposition. As an additional step one may preprocess those mode-$3$ tensors via a \emph{Tucker decomposition}~\cite{KoBa09}
\begin{align}
T_{pqr} = \sum_{\alpha \beta \gamma} R^{(1)}_{p \alpha} R^{(2)}_{q \beta} R^{(3)}_{r \gamma}  S_{\alpha \beta \gamma} \, ,
\end{align}
with factor matrices $R^{(i)}$ and a mode-$3$ \emph{core tensor} $S$. When the Tucker rank $r_\text{Tucker}$ is equal to the original index size $N$ the decomposition becomes exact. Therefore, the Tucker decomposition may enable to perform the CPD on the smaller core tensor $S$, and to subsequently absorb the Tucker factors $R^{(i)}$ into the THC factors. The efficiency of this approach, i.e., the size of the required Tucker ranks needs to be investigated.
\item The present study is, for the sake of simplicity, restricted to a simple second-order perturbative correction to ground-states energies of doubly closed-shell nuclei. In this context, the error on the observable was shown to be one order of magnitude smaller than the error on the input matrix elements. However, no improvement is brought by tensor factorization technique on the computational scaling of such a low-cost many-body calculation. Still, it is to be emphasized that, ultimately, tensor factorization is meant to be applied to more expensive truncation/non-perturbative schemes whose scalings with system size can be convincingly and significantly reduced via tensor factorization. In this context, the tensor network at play in MBPT(2) is a simple, yet representative, example. More involved truncation schemes will yield similar contraction patterns such that, while the size of the associated network is expected to grow, the overall scaling of the tensor-factorized network is not. Therefore, more complicated tensor networks may yield the same complexity as MBPT(2). This field of investigation, which relies on a reformulation of many-body theories in terms of factorized many-body tensors, is expected to attract significant attention in the future. 
\item Antisymmetry of the interaction tensor is presently not enforced during the minimization of the Frobenius norm. Consequently, the reconstructed tensor might not be properly anti-symmetric under permutation of bra and ket indices. This feature relates to a violation of Pauli's exclusion principle. Even though this did not pose apparent problems in the calculation of the second-order energy correction, it was realized in quantum chemistry that enforcing permutation symmetry during the construction of the THC factors is key to converge CC calculations with factorized amplitudes~\cite{Schu17}. Constraining the minimization problem to ensure permutation symmetry is part of our ongoing effort in nuclear theory. 
\end{enumerate}

\section{Conclusions}
\label{sec:conclusion}

In this publication, we have performed the first proof-of-principle applications of tensor-decomposition techniques within the frame of \emph{ab initio} nuclear structure theory. We employ a nuclear Hamiltonian containing two- and three-nucleon interactions derived from chiral effective theory as a testbed for our calculations. In this context, two tensor formats have been explored: the canonical polyadic decomposition and the tensor hypercontraction. In Table~\ref{tab:overview}, an overview of the features displayed by the two tensor formats applied to different tensors is given. While CPD provides a simple approach to compress the tensor the required tensor ranks are quite high. Still the CPD format requires less storage than the original tensor and can be used in subsequent many-body calculations to achieve milder scaling in methods applicable to medium-mass systems. On the other hand the THC format consists of a multi-step procedure where the interaction tensor is pre-processed by compressing the data via a truncated SVD. THC is shown to provide a more adequate tensor format for the nuclear hamiltonian tensor, requiring significantly lower ranks to reach the same accuracy as CPD. This is achieved at the price of needing five instead of four factors matrices, one of which scales quadratically with the THC rank. Both tensor formats are useful in obtaining efficiently compressed nuclear interaction tensors. It is an interesting task for future research to design tensor formats with even better compression properties.

The decomposed interaction tensor is meant to be used in \emph{ab initio} many-body calculations. Second-order MBPT corrections to the ground-state energy of doubly closed-shell nuclei presently provide a reasonable starting point to gauge the error propagated from the tensor-decomposed nuclear interactions to many-body observables. Approximating the matrix elements to an accuracy of $10^{-1}$, the precision on the second-order energy correction is one order of magnitude better.  Such a sub-percent accuracy on the correlation energy is typically reached while reducing the input data by one order of magnitude. While this number is presently obtained in a small single-particle model space, we expect the data compression achieved for a given accuracy on the observable to increase significantly with $e_\text{max}$.

Regarding the tensor decomposition, the next step consists of improving the numerical algorithms in order to allow for applications in larger model spaces. Furthermore, it is envisioned to pre-process left and right singular vectors involved in the THC via a Tucker decomposition. Since the CPD step currently constitutes the limiting factor, this will enable more economical decompositions. As a longer term goal, tensor-decomposition techniques is meant to be extended to genuine three-body forces, i.e., mode-6 tensors. This requires a generalization of the currently employed tensor formats as well as of the underlying numerical algorithms. 

The present work is also meant to be extended to non-perturbative many-body frameworks that, however, need to be reformulated to fully gain from the computational savings of the decompositions. In addition to a significant gain in memory stage, it has the potential to lead to a major reduction of the numerical scaling displayed by high-level many-body methods. In this context, the wish is also to extend the rationale  to representations of the nuclear Hamiltonian in bases appropiate to many-body methods allowing for the breaking (and the restoration) of $U(1)$ and/or $SU(2)$ symmetries; i.e. in quasi-particle and/or $m$-scheme bases, respectively. This is necessary to achieve an efficient \emph{ab initio} description of singly- and doubly-open shell nuclei on the basis of single-reference many-body methods in the future.

Eventually, the long-term hope is that a systematic use of tensor decomposition techniques can help (i) resolve the storage-bottleneck of nuclear three-body matrix elements that currently is one of the main limitations to perform meaningful \emph{ab initio} calculations of nuclei with $A > 100$, and can help (ii) perform highly-accurate many-body calculations at a reduced computational cost. In both cases, the objective is to advance the description and prediction of nuclear phenomena from first principles in conjunction with state-of-the-art experimental studies.

\section*{Acknowledgements}
We thank Robert Roth for providing us with nuclear two- and three-body matrix elements.

This publication is based on work supported in part by the framework of the Espace de Structure et de r\'eactions Nucl\'eaires Th\'eorique (ESNT) at CEA. 
The work at Rice University was supported by the Center for Complex Materials from First Principles, an Energy Frontier Research Center funded by the U.S. Department of Energy, Office of Science, Basic Energy Sciences, under Grant No. DE-SC0012575.

\begin{appendix}

\section{Hartree-Fock theory}
\label{sec:hf}

When treating the 3NF in full in many-body calculations and having the capacity to decompose a mode-6 tensor, it will be of interest to perform the decomposition of the nuclear Hamiltonian tensor in the HO basis. Working with a large enough basis to begin with, one may hope in this way to perform the tensor decomposition only once for a large set of nuclei. Of course, the achieved quality will have to be shown to be maintained independently of the nucleus and the computed observable. 

Let us briefly illustrate this situation  for the THC decomposition of a mode-4 tensor, i.e., for the two-body part of the Hamiltonian. Having at hand the factors $\{Y^{i}_{l_i \alpha}, i\in\{1,2,3,4\}\}$ associated with the tensor-decomposed HO matrix elements and the coefficients of the HF transformation from a prior calculation\footnote{One may envision to perform the HF calculation without any tensor factorization of the Hamiltonian and only build its tensor-factorized form for post-HF calculations.}, Eq.~\eqref{eq:hftrafo} can be rewritten with the help of new factor matrices
\begin{align}
X^{i}_{k_i \alpha} \equiv \sum_{l_i} C_{l_i k_i} Y^{i}_{l_i \alpha} \, ,
\label{eq:factrafo}
\end{align}
under the form
\begin{align}
H^{\text{HF}}_{k_1k_2 k_3 k_4} =\sum_{\alpha \beta} X^1_{k_1 \alpha} X^2_{k_2 \alpha} W_{\alpha \beta} X^3_{k_3 \beta} X^4_{k_4 \beta}\, .
\end{align}
The THC factors of the HF matrix elements are simply those obtained from the HO matrix elements contracted with the HF coefficients. Obviously, the core tensor $W$ is untouched by the transformation. We note that this transformation can be conveniently identified with a simple matrix-matrix product.

In practice, the set of equations must be reformulated in terms of $J$-coupled matrix elements. Given that the transformation coefficients typically arise from a symmetry-restricted, i.e. spherical, HF code, they display the reduced form
\begin{align}
C_{kl} = \delta_{j_k j_l} \delta_{\pi_k \pi_l} \delta_{t_k t_l} C^{(j_k \pi_k t_k)}_{n_k n_l}\, .
\end{align}
Consequently, the HF transformation only mixes radial excitations such that Eq.~\eqref{eq:factrafo} simplifies into
\begin{align}
{^JX}^{i}_{k_i \alpha} = \sum_{n_{l_i}} C_{n_{l_i} n_{k_i}}^{(j_{k_i} \pi_{k_i} t_{k_i})} {^{J\hspace{-1pt}}Y}^{i}_{n_{l_i} \alpha} \, ,
\label{eq:factrafo2}
\end{align}
where ${^J Y}^i$ denotes the THC factors generated from $J$-coupled matrix elements.

\end{appendix}

\bibliographystyle{apsrev4-1}
\bibliography{bib_nucl}

\begin{thebibliography}{49}%
\makeatletter
\providecommand \@ifxundefined [1]{%
 \@ifx{#1\undefined}
}%
\providecommand \@ifnum [1]{%
 \ifnum #1\expandafter \@firstoftwo
 \else \expandafter \@secondoftwo
 \fi
}%
\providecommand \@ifx [1]{%
 \ifx #1\expandafter \@firstoftwo
 \else \expandafter \@secondoftwo
 \fi
}%
\providecommand \natexlab [1]{#1}%
\providecommand \enquote  [1]{``#1''}%
\providecommand \bibnamefont  [1]{#1}%
\providecommand \bibfnamefont [1]{#1}%
\providecommand \citenamefont [1]{#1}%
\providecommand \href@noop [0]{\@secondoftwo}%
\providecommand \href [0]{\begingroup \@sanitize@url \@href}%
\providecommand \@href[1]{\@@startlink{#1}\@@href}%
\providecommand \@@href[1]{\endgroup#1\@@endlink}%
\providecommand \@sanitize@url [0]{\catcode `\\12\catcode `\$12\catcode
  `\&12\catcode `\#12\catcode `\^12\catcode `\_12\catcode `\%12\relax}%
\providecommand \@@startlink[1]{}%
\providecommand \@@endlink[0]{}%
\providecommand \url  [0]{\begingroup\@sanitize@url \@url }%
\providecommand \@url [1]{\endgroup\@href {#1}{\urlprefix }}%
\providecommand \urlprefix  [0]{URL }%
\providecommand \Eprint [0]{\href }%
\providecommand \doibase [0]{http://dx.doi.org/}%
\providecommand \selectlanguage [0]{\@gobble}%
\providecommand \bibinfo  [0]{\@secondoftwo}%
\providecommand \bibfield  [0]{\@secondoftwo}%
\providecommand \translation [1]{[#1]}%
\providecommand \BibitemOpen [0]{}%
\providecommand \bibitemStop [0]{}%
\providecommand \bibitemNoStop [0]{.\EOS\space}%
\providecommand \EOS [0]{\spacefactor3000\relax}%
\providecommand \BibitemShut  [1]{\csname bibitem#1\endcsname}%
\let\auto@bib@innerbib\@empty
\bibitem [{\citenamefont {Kolda}\ and\ \citenamefont {Bader}(2009)}]{KoBa09}%
  \BibitemOpen
  \bibfield  {author} {\bibinfo {author} {\bibfnamefont {T.}~\bibnamefont
  {Kolda}}\ and\ \bibinfo {author} {\bibfnamefont {B.}~\bibnamefont {Bader}},\
  }\href@noop {} {\bibfield  {journal} {\bibinfo  {journal} {SIAM Review}\
  }\textbf {\bibinfo {volume} {51}},\ \bibinfo {pages} {455} (\bibinfo {year}
  {2009})}\BibitemShut {NoStop}%
\bibitem [{\citenamefont {Hohenstein}\ \emph
  {et~al.}(2012{\natexlab{a}})\citenamefont {Hohenstein}, \citenamefont
  {Parrish},\ and\ \citenamefont {Mart{\'\i}nez}}]{Ho12a}%
  \BibitemOpen
  \bibfield  {author} {\bibinfo {author} {\bibfnamefont {E.~G.}\ \bibnamefont
  {Hohenstein}}, \bibinfo {author} {\bibfnamefont {R.~M.}\ \bibnamefont
  {Parrish}}, \ and\ \bibinfo {author} {\bibfnamefont {T.~J.}\ \bibnamefont
  {Mart{\'\i}nez}},\ }\href {\doibase 10.1063/1.4732310} {\bibfield  {journal}
  {\bibinfo  {journal} {The Journal of Chemical Physics}\ }\textbf {\bibinfo
  {volume} {137}},\ \bibinfo {pages} {044103} (\bibinfo {year}
  {2012}{\natexlab{a}})}\BibitemShut {NoStop}%
\bibitem [{\citenamefont {Hohenstein}\ \emph
  {et~al.}(2012{\natexlab{b}})\citenamefont {Hohenstein}, \citenamefont
  {Parrish}, \citenamefont {Sherrill},\ and\ \citenamefont
  {Mart{\'\i}nez}}]{Ho12b}%
  \BibitemOpen
  \bibfield  {author} {\bibinfo {author} {\bibfnamefont {E.~G.}\ \bibnamefont
  {Hohenstein}}, \bibinfo {author} {\bibfnamefont {R.~M.}\ \bibnamefont
  {Parrish}}, \bibinfo {author} {\bibfnamefont {C.~D.}\ \bibnamefont
  {Sherrill}}, \ and\ \bibinfo {author} {\bibfnamefont {T.~J.}\ \bibnamefont
  {Mart{\'\i}nez}},\ }\href {\doibase 10.1063/1.4768241} {\bibfield  {journal}
  {\bibinfo  {journal} {The Journal of Chemical Physics}\ }\textbf {\bibinfo
  {volume} {137}},\ \bibinfo {pages} {221101} (\bibinfo {year}
  {2012}{\natexlab{b}})}\BibitemShut {NoStop}%
\bibitem [{\citenamefont {Parrish}\ \emph {et~al.}(2012)\citenamefont
  {Parrish}, \citenamefont {Hohenstein}, \citenamefont {Mart{\'\i}nez},\ and\
  \citenamefont {Sherrill}}]{Pa12}%
  \BibitemOpen
  \bibfield  {author} {\bibinfo {author} {\bibfnamefont {R.~M.}\ \bibnamefont
  {Parrish}}, \bibinfo {author} {\bibfnamefont {E.~G.}\ \bibnamefont
  {Hohenstein}}, \bibinfo {author} {\bibfnamefont {T.~J.}\ \bibnamefont
  {Mart{\'\i}nez}}, \ and\ \bibinfo {author} {\bibfnamefont {C.~D.}\
  \bibnamefont {Sherrill}},\ }\href {\doibase 10.1063/1.4768233} {\bibfield
  {journal} {\bibinfo  {journal} {The Journal of Chemical Physics}\ }\textbf
  {\bibinfo {volume} {137}},\ \bibinfo {pages} {224106} (\bibinfo {year}
  {2012})}\BibitemShut {NoStop}%
\bibitem [{\citenamefont {Parrish}\ \emph {et~al.}(2013)\citenamefont
  {Parrish}, \citenamefont {Hohenstein}, \citenamefont {Schunck}, \citenamefont
  {Sherrill},\ and\ \citenamefont {Mart\'{\i}nez}}]{Parrish13}%
  \BibitemOpen
  \bibfield  {author} {\bibinfo {author} {\bibfnamefont {R.~M.}\ \bibnamefont
  {Parrish}}, \bibinfo {author} {\bibfnamefont {E.~G.}\ \bibnamefont
  {Hohenstein}}, \bibinfo {author} {\bibfnamefont {N.~F.}\ \bibnamefont
  {Schunck}}, \bibinfo {author} {\bibfnamefont {C.~D.}\ \bibnamefont
  {Sherrill}}, \ and\ \bibinfo {author} {\bibfnamefont {T.~J.}\ \bibnamefont
  {Mart\'{\i}nez}},\ }\href {\doibase 10.1103/PhysRevLett.111.132505}
  {\bibfield  {journal} {\bibinfo  {journal} {Phys. Rev. Lett.}\ }\textbf
  {\bibinfo {volume} {111}},\ \bibinfo {pages} {132505} (\bibinfo {year}
  {2013})}\BibitemShut {NoStop}%
\bibitem [{\citenamefont {Schutski}\ \emph {et~al.}(2017)\citenamefont
  {Schutski}, \citenamefont {Zhao}, \citenamefont {Henderson},\ and\
  \citenamefont {Scuseria}}]{Schu17}%
  \BibitemOpen
  \bibfield  {author} {\bibinfo {author} {\bibfnamefont {R.}~\bibnamefont
  {Schutski}}, \bibinfo {author} {\bibfnamefont {J.}~\bibnamefont {Zhao}},
  \bibinfo {author} {\bibfnamefont {T.~M.}\ \bibnamefont {Henderson}}, \ and\
  \bibinfo {author} {\bibfnamefont {G.~E.}\ \bibnamefont {Scuseria}},\ }\href
  {\doibase 10.1063/1.4996988} {\bibfield  {journal} {\bibinfo  {journal} {The
  Journal of Chemical Physics}\ }\textbf {\bibinfo {volume} {147}},\ \bibinfo
  {pages} {184113} (\bibinfo {year} {2017})}\BibitemShut {NoStop}%
\bibitem [{\citenamefont {Hummel}\ \emph {et~al.}(2017)\citenamefont {Hummel},
  \citenamefont {Tsatsoulis},\ and\ \citenamefont {Gr{\"u}neis}}]{Hu17}%
  \BibitemOpen
  \bibfield  {author} {\bibinfo {author} {\bibfnamefont {F.}~\bibnamefont
  {Hummel}}, \bibinfo {author} {\bibfnamefont {T.}~\bibnamefont {Tsatsoulis}},
  \ and\ \bibinfo {author} {\bibfnamefont {A.}~\bibnamefont {Gr{\"u}neis}},\
  }\href {\doibase 10.1063/1.4977994} {\bibfield  {journal} {\bibinfo
  {journal} {The Journal of Chemical Physics}\ }\textbf {\bibinfo {volume}
  {146}},\ \bibinfo {pages} {124105} (\bibinfo {year} {2017})}\BibitemShut
  {NoStop}%
\bibitem [{\citenamefont {Benedikt}\ \emph {et~al.}(2011)\citenamefont
  {Benedikt}, \citenamefont {Auer}, \citenamefont {Espig},\ and\ \citenamefont
  {Hackbusch}}]{bene11mp2}%
  \BibitemOpen
  \bibfield  {author} {\bibinfo {author} {\bibfnamefont {U.}~\bibnamefont
  {Benedikt}}, \bibinfo {author} {\bibfnamefont {A.~A.}\ \bibnamefont {Auer}},
  \bibinfo {author} {\bibfnamefont {M.}~\bibnamefont {Espig}}, \ and\ \bibinfo
  {author} {\bibfnamefont {W.}~\bibnamefont {Hackbusch}},\ }\href
  {https://doi.org/10.1063/1.3514201} {\bibfield  {journal} {\bibinfo
  {journal} {The Journal of Chemical Physics}\ }\textbf {\bibinfo {volume}
  {134}},\ \bibinfo {pages} {054118} (\bibinfo {year} {2011})}\BibitemShut
  {NoStop}%
\bibitem [{\citenamefont {Benedikt}\ \emph {et~al.}(2013)\citenamefont
  {Benedikt}, \citenamefont {B{\"o}hm},\ and\ \citenamefont
  {Auer}}]{bene13ccd}%
  \BibitemOpen
  \bibfield  {author} {\bibinfo {author} {\bibfnamefont {U.}~\bibnamefont
  {Benedikt}}, \bibinfo {author} {\bibfnamefont {K.-H.}\ \bibnamefont
  {B{\"o}hm}}, \ and\ \bibinfo {author} {\bibfnamefont {A.~A.}\ \bibnamefont
  {Auer}},\ }\href {https://doi.org/10.1063/1.4833565} {\bibfield  {journal}
  {\bibinfo  {journal} {The Journal of Chemical Physics}\ }\textbf {\bibinfo
  {volume} {139}},\ \bibinfo {pages} {224101} (\bibinfo {year}
  {2013})}\BibitemShut {NoStop}%
\bibitem [{\citenamefont {Motta}\ \emph {et~al.}(2018)\citenamefont {Motta},
  \citenamefont {Shee}, \citenamefont {Zhang},\ and\ \citenamefont
  {Chan}}]{MoSh18}%
  \BibitemOpen
  \bibfield  {author} {\bibinfo {author} {\bibfnamefont {M.}~\bibnamefont
  {Motta}}, \bibinfo {author} {\bibfnamefont {J.}~\bibnamefont {Shee}},
  \bibinfo {author} {\bibfnamefont {S.}~\bibnamefont {Zhang}}, \ and\ \bibinfo
  {author} {\bibfnamefont {G.}~\bibnamefont {Chan}},\ }\href@noop {} {\enquote
  {\bibinfo {title} {Efficient ab initio auxiliary-field quantum monte carlo
  calculations in gaussian bases via low-rank tensor decomposition},}\ }
  (\bibinfo {year} {2018}),\ \Eprint {http://arxiv.org/abs/1810.01549}
  {arXiv:1810.01549 [comp-phys]} \BibitemShut {NoStop}%
\bibitem [{\citenamefont {Khoromskaia}\ and\ \citenamefont
  {Khoromskij}(2014)}]{khoromskaia14a}%
  \BibitemOpen
  \bibfield  {author} {\bibinfo {author} {\bibfnamefont {V.}~\bibnamefont
  {Khoromskaia}}\ and\ \bibinfo {author} {\bibfnamefont {B.~N.}\ \bibnamefont
  {Khoromskij}},\ }\href@noop {} {\bibfield  {journal} {\bibinfo  {journal}
  {Comp. Phys. Comm.}\ }\textbf {\bibinfo {volume} {185}},\ \bibinfo {pages}
  {2} (\bibinfo {year} {2014})}\BibitemShut {NoStop}%
\bibitem [{\citenamefont {Lathauwer}\ \emph {et~al.}(2000)\citenamefont
  {Lathauwer}, \citenamefont {Moor},\ and\ \citenamefont
  {Vandewalle}}]{Lathauwer00a}%
  \BibitemOpen
  \bibfield  {author} {\bibinfo {author} {\bibfnamefont {L.~D.}\ \bibnamefont
  {Lathauwer}}, \bibinfo {author} {\bibfnamefont {B.~D.}\ \bibnamefont {Moor}},
  \ and\ \bibinfo {author} {\bibfnamefont {J.}~\bibnamefont {Vandewalle}},\
  }\href@noop {} {\bibfield  {journal} {\bibinfo  {journal} {SIAM J. Mat. Anal.
  Appl.}\ }\textbf {\bibinfo {volume} {21}},\ \bibinfo {pages} {1253} (\bibinfo
  {year} {2000})}\BibitemShut {NoStop}%
\bibitem [{\citenamefont {Som{\`a}}\ \emph {et~al.}(2014)\citenamefont
  {Som{\`a}}, \citenamefont {Cipollone}, \citenamefont {Barbieri},
  \citenamefont {Navr{\'a}til},\ and\ \citenamefont {Duguet}}]{SoCi13}%
  \BibitemOpen
  \bibfield  {author} {\bibinfo {author} {\bibfnamefont {V.}~\bibnamefont
  {Som{\`a}}}, \bibinfo {author} {\bibfnamefont {A.}~\bibnamefont {Cipollone}},
  \bibinfo {author} {\bibfnamefont {C.}~\bibnamefont {Barbieri}}, \bibinfo
  {author} {\bibfnamefont {P.}~\bibnamefont {Navr{\'a}til}}, \ and\ \bibinfo
  {author} {\bibfnamefont {T.}~\bibnamefont {Duguet}},\ }\href@noop {}
  {\bibfield  {journal} {\bibinfo  {journal} {Physical Review C}\ }\textbf
  {\bibinfo {volume} {89}},\ \bibinfo {pages} {061301} (\bibinfo {year}
  {2014})}\BibitemShut {NoStop}%
\bibitem [{\citenamefont {Hergert}\ \emph {et~al.}(2014)\citenamefont
  {Hergert}, \citenamefont {Bogner}, \citenamefont {Morris}, \citenamefont
  {Binder}, \citenamefont {Calci}, \citenamefont {Langhammer},\ and\
  \citenamefont {Roth}}]{Hergert:2014iaa}%
  \BibitemOpen
  \bibfield  {author} {\bibinfo {author} {\bibfnamefont {H.}~\bibnamefont
  {Hergert}}, \bibinfo {author} {\bibfnamefont {S.~K.}\ \bibnamefont {Bogner}},
  \bibinfo {author} {\bibfnamefont {T.~D.}\ \bibnamefont {Morris}}, \bibinfo
  {author} {\bibfnamefont {S.}~\bibnamefont {Binder}}, \bibinfo {author}
  {\bibfnamefont {A.}~\bibnamefont {Calci}}, \bibinfo {author} {\bibfnamefont
  {J.}~\bibnamefont {Langhammer}}, \ and\ \bibinfo {author} {\bibfnamefont
  {R.}~\bibnamefont {Roth}},\ }\href@noop {} {\bibfield  {journal} {\bibinfo
  {journal} {Physical Review C}\ }\textbf {\bibinfo {volume} {90}},\ \bibinfo
  {pages} {041302} (\bibinfo {year} {2014})}\BibitemShut {NoStop}%
\bibitem [{\citenamefont {Jansen}\ \emph {et~al.}(2014)\citenamefont {Jansen},
  \citenamefont {Engel}, \citenamefont {Hagen}, \citenamefont {Navr\'{a}til},\
  and\ \citenamefont {Signoracci}}]{Ja14}%
  \BibitemOpen
  \bibfield  {author} {\bibinfo {author} {\bibfnamefont {G.}~\bibnamefont
  {Jansen}}, \bibinfo {author} {\bibfnamefont {J.}~\bibnamefont {Engel}},
  \bibinfo {author} {\bibfnamefont {G.}~\bibnamefont {Hagen}}, \bibinfo
  {author} {\bibfnamefont {P.}~\bibnamefont {Navr\'{a}til}}, \ and\ \bibinfo
  {author} {\bibfnamefont {A.}~\bibnamefont {Signoracci}},\ }\href {\doibase
  10.1103/PhysRevLett.113.142502} {\bibfield  {journal} {\bibinfo  {journal}
  {Physical Review Letters}\ }\textbf {\bibinfo {volume} {113}},\ \bibinfo
  {pages} {142502} (\bibinfo {year} {2014})}\BibitemShut {NoStop}%
\bibitem [{\citenamefont {Signoracci}\ \emph {et~al.}(2015)\citenamefont
  {Signoracci}, \citenamefont {Duguet}, \citenamefont {Hagen},\ and\
  \citenamefont {Jansen}}]{Si15}%
  \BibitemOpen
  \bibfield  {author} {\bibinfo {author} {\bibfnamefont {A.}~\bibnamefont
  {Signoracci}}, \bibinfo {author} {\bibfnamefont {T.}~\bibnamefont {Duguet}},
  \bibinfo {author} {\bibfnamefont {G.}~\bibnamefont {Hagen}}, \ and\ \bibinfo
  {author} {\bibfnamefont {G.~R.}\ \bibnamefont {Jansen}},\ }\href {\doibase
  10.1103/PhysRevC.91.064320} {\bibfield  {journal} {\bibinfo  {journal}
  {Physical Review C}\ }\textbf {\bibinfo {volume} {91}},\ \bibinfo {pages}
  {064320} (\bibinfo {year} {2015})}\BibitemShut {NoStop}%
\bibitem [{\citenamefont {Bogner}\ \emph {et~al.}(2014)\citenamefont {Bogner},
  \citenamefont {Hergert}, \citenamefont {Holt}, \citenamefont {Schwenk},
  \citenamefont {Binder}, \citenamefont {Calci}, \citenamefont {Langhammer},\
  and\ \citenamefont {Roth}}]{Bo14}%
  \BibitemOpen
  \bibfield  {author} {\bibinfo {author} {\bibfnamefont {S.}~\bibnamefont
  {Bogner}}, \bibinfo {author} {\bibfnamefont {H.}~\bibnamefont {Hergert}},
  \bibinfo {author} {\bibfnamefont {J.~D.}\ \bibnamefont {Holt}}, \bibinfo
  {author} {\bibfnamefont {A.}~\bibnamefont {Schwenk}}, \bibinfo {author}
  {\bibfnamefont {S.}~\bibnamefont {Binder}}, \bibinfo {author} {\bibfnamefont
  {A.}~\bibnamefont {Calci}}, \bibinfo {author} {\bibfnamefont
  {J.}~\bibnamefont {Langhammer}}, \ and\ \bibinfo {author} {\bibfnamefont
  {R.}~\bibnamefont {Roth}},\ }\href {\doibase 10.1103/PhysRevLett.113.142501}
  {\bibfield  {journal} {\bibinfo  {journal} {Physical Review Letters}\
  }\textbf {\bibinfo {volume} {113}},\ \bibinfo {pages} {142501} (\bibinfo
  {year} {2014})}\BibitemShut {NoStop}%
\bibitem [{\citenamefont {Hergert}\ \emph {et~al.}(2016)\citenamefont
  {Hergert}, \citenamefont {Bogner}, \citenamefont {Morris}, \citenamefont
  {Schwenk},\ and\ \citenamefont {Tsukiyama}}]{H15}%
  \BibitemOpen
  \bibfield  {author} {\bibinfo {author} {\bibfnamefont {H.}~\bibnamefont
  {Hergert}}, \bibinfo {author} {\bibfnamefont {S.~K.}\ \bibnamefont {Bogner}},
  \bibinfo {author} {\bibfnamefont {T.~D.}\ \bibnamefont {Morris}}, \bibinfo
  {author} {\bibfnamefont {A.}~\bibnamefont {Schwenk}}, \ and\ \bibinfo
  {author} {\bibfnamefont {K.}~\bibnamefont {Tsukiyama}},\ }\href {\doibase
  10.1016/j.physrep.2015.12.007} {\bibfield  {journal} {\bibinfo  {journal}
  {Physics Reports}\ }\textbf {\bibinfo {volume} {621}},\ \bibinfo {pages}
  {165} (\bibinfo {year} {2016})}\BibitemShut {NoStop}%
\bibitem [{\citenamefont {Gebrerufael}\ \emph {et~al.}(2017)\citenamefont
  {Gebrerufael}, \citenamefont {Vobig}, \citenamefont {Hergert},\ and\
  \citenamefont {Roth}}]{Geb17}%
  \BibitemOpen
  \bibfield  {author} {\bibinfo {author} {\bibfnamefont {E.}~\bibnamefont
  {Gebrerufael}}, \bibinfo {author} {\bibfnamefont {K.}~\bibnamefont {Vobig}},
  \bibinfo {author} {\bibfnamefont {H.}~\bibnamefont {Hergert}}, \ and\
  \bibinfo {author} {\bibfnamefont {R.}~\bibnamefont {Roth}},\ }\href {\doibase
  10.1103/PhysRevLett.118.152503} {\bibfield  {journal} {\bibinfo  {journal}
  {Physical Review Letters}\ }\textbf {\bibinfo {volume} {118}},\ \bibinfo
  {pages} {152503} (\bibinfo {year} {2017})}\BibitemShut {NoStop}%
\bibitem [{\citenamefont {Tichai}\ \emph
  {et~al.}(2018{\natexlab{a}})\citenamefont {Tichai}, \citenamefont
  {Gebrerufael},\ and\ \citenamefont {Roth}}]{Ti18a}%
  \BibitemOpen
  \bibfield  {author} {\bibinfo {author} {\bibfnamefont {A.}~\bibnamefont
  {Tichai}}, \bibinfo {author} {\bibfnamefont {E.}~\bibnamefont {Gebrerufael}},
  \ and\ \bibinfo {author} {\bibfnamefont {R.}~\bibnamefont {Roth}},\
  }\href@noop {} {\enquote {\bibinfo {title} {Open-shell nuclei from no-core
  shell model with perturbative improvement},}\ } (\bibinfo {year}
  {2018}{\natexlab{a}}),\ \bibinfo {note} {(accepted for publication in Physics
  Letters B)},\ \Eprint {http://arxiv.org/abs/1703.05664} {arXiv:1703.05664
  [nucl-th]} \BibitemShut {NoStop}%
\bibitem [{\citenamefont {Tichai}\ \emph
  {et~al.}(2018{\natexlab{b}})\citenamefont {Tichai}, \citenamefont {Arthuis},
  \citenamefont {Duguet}, \citenamefont {Hergert}, \citenamefont {Som{\`a}},\
  and\ \citenamefont {Roth}}]{Ti18b}%
  \BibitemOpen
  \bibfield  {author} {\bibinfo {author} {\bibfnamefont {A.}~\bibnamefont
  {Tichai}}, \bibinfo {author} {\bibfnamefont {P.}~\bibnamefont {Arthuis}},
  \bibinfo {author} {\bibfnamefont {T.}~\bibnamefont {Duguet}}, \bibinfo
  {author} {\bibfnamefont {H.}~\bibnamefont {Hergert}}, \bibinfo {author}
  {\bibfnamefont {V.}~\bibnamefont {Som{\`a}}}, \ and\ \bibinfo {author}
  {\bibfnamefont {R.}~\bibnamefont {Roth}},\ }\href@noop {} {\bibfield
  {journal} {\bibinfo  {journal} {Phys. Lett. B}\ }\textbf {\bibinfo {volume}
  {786}},\ \bibinfo {pages} {195} (\bibinfo {year}
  {2018}{\natexlab{b}})}\BibitemShut {NoStop}%
\bibitem [{\citenamefont {Arthuis}\ \emph {et~al.}(2018)\citenamefont
  {Arthuis}, \citenamefont {Duguet}, \citenamefont {Tichai}, \citenamefont
  {Lasseri},\ and\ \citenamefont {Ebran}}]{Art18a}%
  \BibitemOpen
  \bibfield  {author} {\bibinfo {author} {\bibfnamefont {P.}~\bibnamefont
  {Arthuis}}, \bibinfo {author} {\bibfnamefont {T.}~\bibnamefont {Duguet}},
  \bibinfo {author} {\bibfnamefont {A.}~\bibnamefont {Tichai}}, \bibinfo
  {author} {\bibfnamefont {R.-D.}\ \bibnamefont {Lasseri}}, \ and\ \bibinfo
  {author} {\bibfnamefont {J.-P.}\ \bibnamefont {Ebran}},\ }\href@noop {} {\
  (\bibinfo {year} {2018})},\ \Eprint {http://arxiv.org/abs/1809.01187}
  {arXiv:1809.01187 [nucl-th]} \BibitemShut {NoStop}%
\bibitem [{\citenamefont {Duguet}(2015)}]{Du15}%
  \BibitemOpen
  \bibfield  {author} {\bibinfo {author} {\bibfnamefont {T.}~\bibnamefont
  {Duguet}},\ }\href {\doibase 10.1088/0954-3899/42/2/025107} {\bibfield
  {journal} {\bibinfo  {journal} {Journal of Physics G: Nuclear and Particle
  Physics}\ }\textbf {\bibinfo {volume} {42}},\ \bibinfo {pages} {025107}
  (\bibinfo {year} {2015})}\BibitemShut {NoStop}%
\bibitem [{\citenamefont {Duguet}\ and\ \citenamefont
  {Signoracci}(2016)}]{Du16}%
  \BibitemOpen
  \bibfield  {author} {\bibinfo {author} {\bibfnamefont {T.}~\bibnamefont
  {Duguet}}\ and\ \bibinfo {author} {\bibfnamefont {A.}~\bibnamefont
  {Signoracci}},\ }\href@noop {} {\bibfield  {journal} {\bibinfo  {journal}
  {Journal of Physics G: Nuclear and Particle Physics}\ }\textbf {\bibinfo
  {volume} {44}} (\bibinfo {year} {2016})}\BibitemShut {NoStop}%
\bibitem [{\citenamefont {Qiu}\ \emph {et~al.}(2017)\citenamefont {Qiu},
  \citenamefont {Henderson}, \citenamefont {Zhao},\ and\ \citenamefont
  {Scuseria}}]{Qiu17}%
  \BibitemOpen
  \bibfield  {author} {\bibinfo {author} {\bibfnamefont {Y.}~\bibnamefont
  {Qiu}}, \bibinfo {author} {\bibfnamefont {T.~M.}\ \bibnamefont {Henderson}},
  \bibinfo {author} {\bibfnamefont {J.}~\bibnamefont {Zhao}}, \ and\ \bibinfo
  {author} {\bibfnamefont {G.~E.}\ \bibnamefont {Scuseria}},\ }\href@noop {}
  {\bibfield  {journal} {\bibinfo  {journal} {The Journal of Chemical Physics}\
  }\textbf {\bibinfo {volume} {147}} (\bibinfo {year} {2017})}\BibitemShut
  {NoStop}%
\bibitem [{\citenamefont {Hergert}(2017)}]{Hergert:2017kx}%
  \BibitemOpen
  \bibfield  {author} {\bibinfo {author} {\bibfnamefont {H.}~\bibnamefont
  {Hergert}},\ }\href@noop {} {\bibfield  {journal} {\bibinfo  {journal} {Phys.
  Scripta}\ }\textbf {\bibinfo {volume} {92}},\ \bibinfo {pages} {023002}
  (\bibinfo {year} {2017})}\BibitemShut {NoStop}%
\bibitem [{\citenamefont {Dickhoff}\ and\ \citenamefont
  {Barbieri}(2004)}]{Dickhoff:2004xx}%
  \BibitemOpen
  \bibfield  {author} {\bibinfo {author} {\bibfnamefont {W.~H.}\ \bibnamefont
  {Dickhoff}}\ and\ \bibinfo {author} {\bibfnamefont {C.}~\bibnamefont
  {Barbieri}},\ }\href@noop {} {\bibfield  {journal} {\bibinfo  {journal}
  {Prog. Part. Nucl. Phys.}\ }\textbf {\bibinfo {volume} {52}},\ \bibinfo
  {pages} {377} (\bibinfo {year} {2004})}\BibitemShut {NoStop}%
\bibitem [{\citenamefont {Raimondi}\ and\ \citenamefont
  {Barbieri}(2018)}]{Raimondi:2018aa}%
  \BibitemOpen
  \bibfield  {author} {\bibinfo {author} {\bibfnamefont {F.}~\bibnamefont
  {Raimondi}}\ and\ \bibinfo {author} {\bibfnamefont {C.}~\bibnamefont
  {Barbieri}},\ }\href {\doibase 10.1103/PhysRevC.97.054308} {\bibfield
  {journal} {\bibinfo  {journal} {Physical Review C}\ }\textbf {\bibinfo
  {volume} {97}},\ \bibinfo {pages} {054308} (\bibinfo {year}
  {2018})}\BibitemShut {NoStop}%
\bibitem [{\citenamefont {Vervliet}\ \emph {et~al.}()\citenamefont {Vervliet},
  \citenamefont {Debals}, \citenamefont {Sorber}, \citenamefont {Van~Barel},\
  and\ \citenamefont {De~Lathauwer}}]{tensorlab3.0}%
  \BibitemOpen
  \bibfield  {author} {\bibinfo {author} {\bibfnamefont {N.}~\bibnamefont
  {Vervliet}}, \bibinfo {author} {\bibfnamefont {O.}~\bibnamefont {Debals}},
  \bibinfo {author} {\bibfnamefont {L.}~\bibnamefont {Sorber}}, \bibinfo
  {author} {\bibfnamefont {M.}~\bibnamefont {Van~Barel}}, \ and\ \bibinfo
  {author} {\bibfnamefont {L.}~\bibnamefont {De~Lathauwer}},\ }\href@noop {}
  {\enquote {\bibinfo {title} {Tensorlab 3.0},}\ }\BibitemShut {NoStop}%
\bibitem [{\citenamefont {Bogner}\ \emph {et~al.}(2007)\citenamefont {Bogner},
  \citenamefont {Furnstahl},\ and\ \citenamefont {Perry}}]{BoFu07}%
  \BibitemOpen
  \bibfield  {author} {\bibinfo {author} {\bibfnamefont {S.~K.}\ \bibnamefont
  {Bogner}}, \bibinfo {author} {\bibfnamefont {R.~J.}\ \bibnamefont
  {Furnstahl}}, \ and\ \bibinfo {author} {\bibfnamefont {R.~J.}\ \bibnamefont
  {Perry}},\ }\href@noop {} {\bibfield  {journal} {\bibinfo  {journal}
  {Physical Review C}\ }\textbf {\bibinfo {volume} {75}},\ \bibinfo {pages}
  {061001(R)} (\bibinfo {year} {2007})}\BibitemShut {NoStop}%
\bibitem [{\citenamefont {Hergert}\ and\ \citenamefont {Roth}(2007)}]{HeRo07}%
  \BibitemOpen
  \bibfield  {author} {\bibinfo {author} {\bibfnamefont {H.}~\bibnamefont
  {Hergert}}\ and\ \bibinfo {author} {\bibfnamefont {R.}~\bibnamefont {Roth}},\
  }\href@noop {} {\bibfield  {journal} {\bibinfo  {journal} {Physical Review
  C}\ }\textbf {\bibinfo {volume} {75}},\ \bibinfo {pages} {051001(R)}
  (\bibinfo {year} {2007})}\BibitemShut {NoStop}%
\bibitem [{\citenamefont {Roth}\ \emph {et~al.}(2008)\citenamefont {Roth},
  \citenamefont {Reinhardt},\ and\ \citenamefont {Hergert}}]{RoRe08}%
  \BibitemOpen
  \bibfield  {author} {\bibinfo {author} {\bibfnamefont {R.}~\bibnamefont
  {Roth}}, \bibinfo {author} {\bibfnamefont {S.}~\bibnamefont {Reinhardt}}, \
  and\ \bibinfo {author} {\bibfnamefont {H.}~\bibnamefont {Hergert}},\
  }\href@noop {} {\bibfield  {journal} {\bibinfo  {journal} {Physical Review
  C}\ }\textbf {\bibinfo {volume} {77}},\ \bibinfo {pages} {064003} (\bibinfo
  {year} {2008})}\BibitemShut {NoStop}%
\bibitem [{\citenamefont {Roth}\ \emph {et~al.}(2011)\citenamefont {Roth},
  \citenamefont {Langhammer}, \citenamefont {Calci}, \citenamefont {Binder},\
  and\ \citenamefont {Navr{\'a}til}}]{RoLa11}%
  \BibitemOpen
  \bibfield  {author} {\bibinfo {author} {\bibfnamefont {R.}~\bibnamefont
  {Roth}}, \bibinfo {author} {\bibfnamefont {J.}~\bibnamefont {Langhammer}},
  \bibinfo {author} {\bibfnamefont {A.}~\bibnamefont {Calci}}, \bibinfo
  {author} {\bibfnamefont {S.}~\bibnamefont {Binder}}, \ and\ \bibinfo {author}
  {\bibfnamefont {P.}~\bibnamefont {Navr{\'a}til}},\ }\href {\doibase
  10.1103/PhysRevLett.107.072501} {\bibfield  {journal} {\bibinfo  {journal}
  {Physical Review Letters}\ }\textbf {\bibinfo {volume} {107}},\ \bibinfo
  {pages} {072501} (\bibinfo {year} {2011})}\BibitemShut {NoStop}%
\bibitem [{\citenamefont {Jurgenson}\ \emph {et~al.}(2013)\citenamefont
  {Jurgenson}, \citenamefont {Maris}, \citenamefont {Furnstahl}, \citenamefont
  {Navr{\'a}til}, \citenamefont {Ormand},\ and\ \citenamefont {Vary}}]{JuMa13}%
  \BibitemOpen
  \bibfield  {author} {\bibinfo {author} {\bibfnamefont {E.~D.}\ \bibnamefont
  {Jurgenson}}, \bibinfo {author} {\bibfnamefont {P.}~\bibnamefont {Maris}},
  \bibinfo {author} {\bibfnamefont {R.~J.}\ \bibnamefont {Furnstahl}}, \bibinfo
  {author} {\bibfnamefont {P.}~\bibnamefont {Navr{\'a}til}}, \bibinfo {author}
  {\bibfnamefont {W.~E.}\ \bibnamefont {Ormand}}, \ and\ \bibinfo {author}
  {\bibfnamefont {J.~P.}\ \bibnamefont {Vary}},\ }\href@noop {} {\bibfield
  {journal} {\bibinfo  {journal} {Physical Review C}\ }\textbf {\bibinfo
  {volume} {87}},\ \bibinfo {pages} {054312} (\bibinfo {year}
  {2013})}\BibitemShut {NoStop}%
\bibitem [{\citenamefont {Entem}\ and\ \citenamefont
  {Machleidt}(2003)}]{EnMa03}%
  \BibitemOpen
  \bibfield  {author} {\bibinfo {author} {\bibfnamefont {D.~R.}\ \bibnamefont
  {Entem}}\ and\ \bibinfo {author} {\bibfnamefont {R.}~\bibnamefont
  {Machleidt}},\ }\href@noop {} {\bibfield  {journal} {\bibinfo  {journal}
  {Physical Review C}\ }\textbf {\bibinfo {volume} {68}},\ \bibinfo {pages}
  {041001(R)} (\bibinfo {year} {2003})}\BibitemShut {NoStop}%
\bibitem [{\citenamefont {Navr\'atil}(2007)}]{Na07}%
  \BibitemOpen
  \bibfield  {author} {\bibinfo {author} {\bibfnamefont {P.}~\bibnamefont
  {Navr\'atil}},\ }\href {\doibase 10.1007/s00601-007-0193-3} {\bibfield
  {journal} {\bibinfo  {journal} {Few Body Systems}\ }\textbf {\bibinfo
  {volume} {41}},\ \bibinfo {pages} {117} (\bibinfo {year} {2007})}\BibitemShut
  {NoStop}%
\bibitem [{\citenamefont {Roth}\ \emph
  {et~al.}(2012{\natexlab{a}})\citenamefont {Roth}, \citenamefont {Binder},
  \citenamefont {Vobig}, \citenamefont {Calci}, \citenamefont {Langhammer},\
  and\ \citenamefont {Navratil}}]{Roth:2011vt}%
  \BibitemOpen
  \bibfield  {author} {\bibinfo {author} {\bibfnamefont {R.}~\bibnamefont
  {Roth}}, \bibinfo {author} {\bibfnamefont {S.}~\bibnamefont {Binder}},
  \bibinfo {author} {\bibfnamefont {K.}~\bibnamefont {Vobig}}, \bibinfo
  {author} {\bibfnamefont {A.}~\bibnamefont {Calci}}, \bibinfo {author}
  {\bibfnamefont {J.}~\bibnamefont {Langhammer}}, \ and\ \bibinfo {author}
  {\bibfnamefont {P.}~\bibnamefont {Navratil}},\ }\href {\doibase
  10.1103/PhysRevLett.109.052501} {\bibfield  {journal} {\bibinfo  {journal}
  {Physical Review Letters}\ }\textbf {\bibinfo {volume} {109}},\ \bibinfo
  {pages} {052501} (\bibinfo {year} {2012}{\natexlab{a}})}\BibitemShut
  {NoStop}%
\bibitem [{\citenamefont {Hergert}\ \emph {et~al.}(2013)\citenamefont
  {Hergert}, \citenamefont {Binder}, \citenamefont {Calci}, \citenamefont
  {Langhammer},\ and\ \citenamefont {Roth}}]{HeBi13}%
  \BibitemOpen
  \bibfield  {author} {\bibinfo {author} {\bibfnamefont {H.}~\bibnamefont
  {Hergert}}, \bibinfo {author} {\bibfnamefont {S.}~\bibnamefont {Binder}},
  \bibinfo {author} {\bibfnamefont {A.}~\bibnamefont {Calci}}, \bibinfo
  {author} {\bibfnamefont {J.}~\bibnamefont {Langhammer}}, \ and\ \bibinfo
  {author} {\bibfnamefont {R.}~\bibnamefont {Roth}},\ }\href@noop {} {\bibfield
   {journal} {\bibinfo  {journal} {Physical Review Letters}\ }\textbf {\bibinfo
  {volume} {110}},\ \bibinfo {pages} {242501} (\bibinfo {year}
  {2013})}\BibitemShut {NoStop}%
\bibitem [{\citenamefont {Binder}\ \emph {et~al.}(2013)\citenamefont {Binder},
  \citenamefont {Langhammer}, \citenamefont {Calci}, \citenamefont
  {Navr{\'a}til},\ and\ \citenamefont {Roth}}]{BiLa13}%
  \BibitemOpen
  \bibfield  {author} {\bibinfo {author} {\bibfnamefont {S.}~\bibnamefont
  {Binder}}, \bibinfo {author} {\bibfnamefont {J.}~\bibnamefont {Langhammer}},
  \bibinfo {author} {\bibfnamefont {A.}~\bibnamefont {Calci}}, \bibinfo
  {author} {\bibfnamefont {P.}~\bibnamefont {Navr{\'a}til}}, \ and\ \bibinfo
  {author} {\bibfnamefont {R.}~\bibnamefont {Roth}},\ }\href@noop {} {\bibfield
   {journal} {\bibinfo  {journal} {Physical Review C}\ }\textbf {\bibinfo
  {volume} {87}},\ \bibinfo {pages} {021303} (\bibinfo {year}
  {2013})}\BibitemShut {NoStop}%
\bibitem [{\citenamefont {Tichai}\ \emph {et~al.}(2016)\citenamefont {Tichai},
  \citenamefont {Langhammer}, \citenamefont {Binder},\ and\ \citenamefont
  {Roth}}]{Ti16}%
  \BibitemOpen
  \bibfield  {author} {\bibinfo {author} {\bibfnamefont {A.}~\bibnamefont
  {Tichai}}, \bibinfo {author} {\bibfnamefont {J.}~\bibnamefont {Langhammer}},
  \bibinfo {author} {\bibfnamefont {S.}~\bibnamefont {Binder}}, \ and\ \bibinfo
  {author} {\bibfnamefont {R.}~\bibnamefont {Roth}},\ }\href {\doibase
  10.1016/j.physletb.2016.03.029} {\bibfield  {journal} {\bibinfo  {journal}
  {Physics Letters B}\ }\textbf {\bibinfo {volume} {756}},\ \bibinfo {pages}
  {283 } (\bibinfo {year} {2016})}\BibitemShut {NoStop}%
\bibitem [{\citenamefont {Hu}\ \emph {et~al.}(2016)\citenamefont {Hu},
  \citenamefont {Xu}, \citenamefont {Sun}, \citenamefont {Vary},\ and\
  \citenamefont {Li}}]{Hu16}%
  \BibitemOpen
  \bibfield  {author} {\bibinfo {author} {\bibfnamefont {B.~S.}\ \bibnamefont
  {Hu}}, \bibinfo {author} {\bibfnamefont {F.~R.}\ \bibnamefont {Xu}}, \bibinfo
  {author} {\bibfnamefont {Z.~H.}\ \bibnamefont {Sun}}, \bibinfo {author}
  {\bibfnamefont {J.~P.}\ \bibnamefont {Vary}}, \ and\ \bibinfo {author}
  {\bibfnamefont {T.}~\bibnamefont {Li}},\ }\href@noop {} {\bibfield  {journal}
  {\bibinfo  {journal} {Physical Review C}\ }\textbf {\bibinfo {volume} {94}}
  (\bibinfo {year} {2016})}\BibitemShut {NoStop}%
\bibitem [{\citenamefont {Roth}\ \emph
  {et~al.}(2012{\natexlab{b}})\citenamefont {Roth}, \citenamefont {Binder},
  \citenamefont {Vobig}, \citenamefont {Calci}, \citenamefont {Langhammer},\
  and\ \citenamefont {Navr\'atil}}]{RoBi12}%
  \BibitemOpen
  \bibfield  {author} {\bibinfo {author} {\bibfnamefont {R.}~\bibnamefont
  {Roth}}, \bibinfo {author} {\bibfnamefont {S.}~\bibnamefont {Binder}},
  \bibinfo {author} {\bibfnamefont {K.}~\bibnamefont {Vobig}}, \bibinfo
  {author} {\bibfnamefont {A.}~\bibnamefont {Calci}}, \bibinfo {author}
  {\bibfnamefont {J.}~\bibnamefont {Langhammer}}, \ and\ \bibinfo {author}
  {\bibfnamefont {P.}~\bibnamefont {Navr\'atil}},\ }\href@noop {} {\bibfield
  {journal} {\bibinfo  {journal} {Physical Review Letters}\ }\textbf {\bibinfo
  {volume} {109}},\ \bibinfo {pages} {052501} (\bibinfo {year}
  {2012}{\natexlab{b}})}\BibitemShut {NoStop}%
\bibitem [{\citenamefont {Gebrerufael}\ \emph {et~al.}(2016)\citenamefont
  {Gebrerufael}, \citenamefont {Calci},\ and\ \citenamefont {Roth}}]{Geb16}%
  \BibitemOpen
  \bibfield  {author} {\bibinfo {author} {\bibfnamefont {E.}~\bibnamefont
  {Gebrerufael}}, \bibinfo {author} {\bibfnamefont {A.}~\bibnamefont {Calci}},
  \ and\ \bibinfo {author} {\bibfnamefont {R.}~\bibnamefont {Roth}},\
  }\href@noop {} {\bibfield  {journal} {\bibinfo  {journal} {Physical Review
  C}\ }\textbf {\bibinfo {volume} {93}} (\bibinfo {year} {2016})}\BibitemShut
  {NoStop}%
\bibitem [{\citenamefont {Szabo}\ and\ \citenamefont {Ostlund}(1982)}]{SzOs82}%
  \BibitemOpen
  \bibfield  {author} {\bibinfo {author} {\bibfnamefont {A.}~\bibnamefont
  {Szabo}}\ and\ \bibinfo {author} {\bibfnamefont {N.~S.}\ \bibnamefont
  {Ostlund}},\ }\href@noop {} {\emph {\bibinfo {title} {Modern Quantum
  Chemistry}}}\ (\bibinfo  {publisher} {Dover Publications Inc.},\ \bibinfo
  {year} {1982})\BibitemShut {NoStop}%
\bibitem [{\citenamefont {Shavitt}\ and\ \citenamefont
  {Bartlett}(2009)}]{ShBa09}%
  \BibitemOpen
  \bibfield  {author} {\bibinfo {author} {\bibfnamefont {I.}~\bibnamefont
  {Shavitt}}\ and\ \bibinfo {author} {\bibfnamefont {R.~J.}\ \bibnamefont
  {Bartlett}},\ }\href@noop {} {\emph {\bibinfo {title} {Many-body methods in
  chemistry and physics}}}\ (\bibinfo  {publisher} {Cambridge University
  Press},\ \bibinfo {year} {2009})\BibitemShut {NoStop}%
\bibitem [{\citenamefont {Varshalovich}\ \emph {et~al.}(1988)\citenamefont
  {Varshalovich}, \citenamefont {Moskalev},\ and\ \citenamefont
  {Khersonskii}}]{VaMo88}%
  \BibitemOpen
  \bibfield  {author} {\bibinfo {author} {\bibfnamefont {D.~A.}\ \bibnamefont
  {Varshalovich}}, \bibinfo {author} {\bibfnamefont {A.~N.}\ \bibnamefont
  {Moskalev}}, \ and\ \bibinfo {author} {\bibfnamefont {V.~K.}\ \bibnamefont
  {Khersonskii}},\ }\href@noop {} {\emph {\bibinfo {title} {Quantum Theory of
  Angular Momentum}}}\ (\bibinfo  {publisher} {World Scientific Publishing
  Company},\ \bibinfo {year} {1988})\BibitemShut {NoStop}%
\bibitem [{\citenamefont {Braess}\ and\ \citenamefont
  {Hackbusch}(2005)}]{BrHa05}%
  \BibitemOpen
  \bibfield  {author} {\bibinfo {author} {\bibfnamefont {D.}~\bibnamefont
  {Braess}}\ and\ \bibinfo {author} {\bibfnamefont {W.}~\bibnamefont
  {Hackbusch}},\ }\href {\doibase 10.1093/imanum/dri015} {\bibfield  {journal}
  {\bibinfo  {journal} {IMA Journal of Numerical Analysis}\ }\textbf {\bibinfo
  {volume} {25}},\ \bibinfo {pages} {685} (\bibinfo {year} {2005})}\BibitemShut
  {NoStop}%
\bibitem [{\citenamefont {Suhonen}(2007)}]{Su07}%
  \BibitemOpen
  \bibfield  {author} {\bibinfo {author} {\bibfnamefont {J.}~\bibnamefont
  {Suhonen}},\ }\href {\doibase 10.1007/978-3-540-48861-3} {\emph {\bibinfo
  {title} {{From Nucleons to Nucleus}}}},\ Theoretical and Mathematical
  Physics\ (\bibinfo  {publisher} {Springer},\ \bibinfo {address} {Berlin,
  Germany},\ \bibinfo {year} {2007})\BibitemShut {NoStop}%
\bibitem [{\citenamefont {Tichai}\ and\ \citenamefont {Duguet}(2018)}]{TiDu18}%
  \BibitemOpen
  \bibfield  {author} {\bibinfo {author} {\bibfnamefont {A.}~\bibnamefont
  {Tichai}}\ and\ \bibinfo {author} {\bibfnamefont {T.}~\bibnamefont
  {Duguet}},\ }\href@noop {} {} (\bibinfo {year} {2018}),\ \bibinfo {note}
  {(unpublished)}\BibitemShut {NoStop}%
\end{thebibliography}%

\end{document}